\documentclass[useAMS,referee]{article}
\usepackage[utf8]{inputenc}
\usepackage{graphicx}
\usepackage{multirow}
\usepackage[left=1.5cm,right=1.5cm,top=1.5cm,bottom=1.5cm]{geometry}
\usepackage{url}
\usepackage{mathrsfs}
\usepackage[figuresright]{rotating}
\usepackage{latexsym,amssymb,amsfonts,amsmath,dsfont,enumerate,color}

\newcommand{\R}{\mathbb{R}}

\newcommand{\I}{\mathds{1}}
\def\vec{\boldsymbol}
\def\bSig\mathbf{\Sigma}

\def\vec{\boldsymbol}

\title{\textbf{\LARGE{Modeling excess hazard with time--to--cure as a parameter}}}
\author{\textbf{Olayid\'e Boussari}$^{1,2,*}$, \textbf{Laurent Bordes}$^{3}$, \textbf{Ga\"elle Romain}$^{1,2}$, \textbf{Marc Colonna}$^{4}$,\\ \textbf{Nadine Bossard}$^{5,6}$, \textbf{Laurent Remontet}$^{5,6}$, and \textbf{Val\'erie Jooste}$^{1,2}$ 
\vspace{0.4cm} \\ 
$^{1}$\small{UMR 1231, EPICAD team, INSERM, Universit\'e Bourgogne--Franche--Comt\'e, Dijon, F--21000, France}\\
$^{2}$\small{Registre Bourguignon des Cancers Digestifs, Dijon-Bourgogne University Hospital, Dijon, F--21000, France} \\
$^{3}$\small{UMR 5142, LMAP--IPRA, CNRS, E2S UPPA, Universit\'e Pau \& Pays Adour, Pau, F--64000, France}   \\
$^{4}$\small{Registre du Cancer de l'Is\`ere, Grenoble University Hospital, Grenoble, F--38000, France} \\
$^{5}$\small{Department of Biostatistics and Bioinformatics, Hospices Civils de Lyon, Lyon, F--69003, France} \\
$^{6}$\small{UMR 5558, LBBE, Biostatistics Health Group, CNRS, University Lyon 1, Lyon, F--69100, France} \\
$^{*}$\small{\textit{Email}: olayide.boussari@u-bourgogne.fr} }

\date{} 

\begin{document}
\maketitle
\vspace{1cm}
\begin{abstract}
\linespread{1.2}\selectfont
\noindent Cure models have been widely developed to estimate the cure fraction when some subjects never experience the event of interest. However these models were rarely focused on the estimation of the time--to--cure i.e. the delay elapsed between the diagnosis and "the time from which cure is reached", an important indicator, for instance to address  the question of access to insurance or loans for subjects with personal history of cancer. We propose a new excess hazard regression model that includes the time--to--cure as a covariate dependent parameter to be estimated. The model is written similarly to a Beta probability distribution function and is shown to be a particular case of the non-mixture cure models. Parameters are estimated through a maximum likelihood approach and simulation studies demonstrate good performance of the model. Illustrative applications to two cancer data sets are provided and some limitations as well as possible extensions of the model are discussed. The proposed model offers a simple and comprehensive way to estimate more accurately the time--to--cure.\\

\end{abstract}

\newpage

\section{Background} \label{S1}
\begin{figure}
\centerline{\includegraphics[width=16cm,height=12cm]{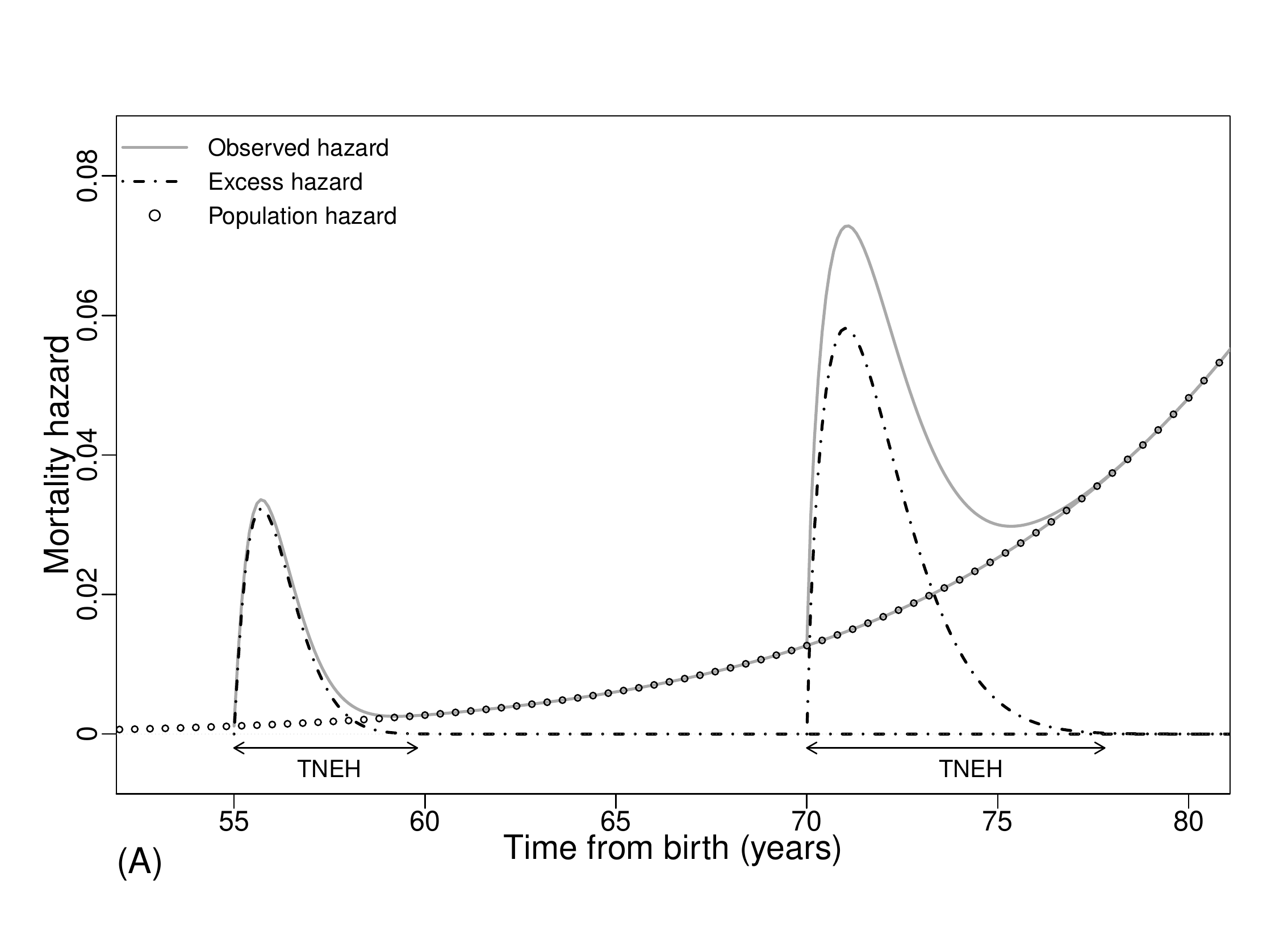}}
\caption{Illustrative plot of Observed, Excess and Population hazards functions for an individual diagnosed at $55$, $70$ years.}
\label{fig1}
\end{figure}

Since their first formulation by Boag (1949), Mixture Cure Models (MCM) have been widely developed to deal with survival data including a fraction of subjects who never experience the event of interest ("cured subjects"). Various ways have been considered by authors (see e.g. Kuk and Chen (1992); Li and Taylor (2002); Zhang and Peng (2009)) to model the baseline of both  the survival function of the "uncured subjects" and the cure fraction, as well as the covariates effects on these two quantities, and these led to an extensive development of the MCM. A large review of MCM can be found in Maller and Zhou (1996) or Klein et al. (2016). 
In Yakovlev, Tsodikov, and Asselain (1996) a new family of cure models was introduced, the bounded cumulative hazard cure models also known as non-mixture cure models (NMCM). Suitable reviews and interpretations of the NMCM were proposed in Chen, Ibrahim, and Sinha (1999); Tsodikov, Ibrahim, and Yakovlev (2003); and Cooner et al. (2007). 
Other cure models can be found in the litterature (e.g. Yin and Ibrahim (2005b); Gu, Sinha, and Banerjee (2011)) and some approaches based on the Box-Cox transformation have been developed to  unify different types of cure models (Yin and Ibrahim (2005a); Zeng, Yin, and Ibrahim (2006); Taylor and Liu (2007)). For practical use and interpretation of covariates effects, each type of cure model has both advantages and disadvantages and the issue of cure model selection was addressed for example in Peng and Xu (2012).\\
Since the late 1990s, cure models have been extended to the framework of the net survival (survival that would be observed if no death could occur from other causes than the disease of interest) with applications emphasized on cancer data (see Verdecchia et al. (1998); Yu et al. (2005); Lambert et al. (2006); Andersson et al. (2011) among others). \\
Let us recall briefly hereafter the basic concept of cure models within the net survival framework. We consider a population of patients suffering from a disease (say cancer). For a given patient, let denote $A$ the age at diagnosis, $X_{1}$ ($X_{2}$ respectively) the latent variable corresponding to the time elapsed between the diagnosis and the death due to cancer (other causes respectively), $C$ the right censoring time, $\Delta=\I\{{X<C}\}$ the censoring indicator where $X=\min(X_{1},X_{2})$ and $\vec Z$ a vector of covariates in $\R^d$. We assume that conditionally on $\vec Z$, $X$ and $C$ are independent.
Then an observation is a quadruple $(T,\Delta,A,\vec Z)$, where $T=\min(X,C)$ is the observed time since diagnosis then $T+A$ is the observed time since birth. As it is well known that the age at diagnosis is one of the covariates that influenced the risk of dying from cancer, $A$ is often also included in the vector of covariates $\vec Z$. When cause of death is not available, the most used methodology to estimate the net survival is to assume that conditionally on $(A,\vec Z)=(a,\vec z)$, the observed hazard $\lambda_{\rm obs}$ of $T$ equals the sum of $\lambda_{\rm pop}$ the known background mortality hazard in the general population (provided by life tables from national statistics) and $\lambda_{\rm exc}$ the excess hazard due to cancer:
\begin{equation} 
\lambda_{\rm obs}(t|\vec z)=\lambda_{\rm pop}(t+a|\vec z) + \lambda_{\rm exc}(t|\vec z).
\label{eq1} 
\end{equation}
From~\eqref{eq1} the link between the survival functions is given by:
\begin{equation} 
S_{\rm obs}(t|\vec z)=S_{\rm pop}(t+a|\vec z) \times S_{\rm net}(t|\vec z),
\label{eq2}
\end{equation}
where $S_{\rm obs}$ ($S_{\rm pop}$ respectively) represents the observed (the background respectively) survival distribution function, and $S_{\rm net}$ corresponds to the net survival distribution function, linked to $\lambda_{\rm exc}$ through 
\begin{equation} 
S_{\rm net}(t|\vec z)=\exp\Bigl\{-\Lambda_{\rm exc}(t|\vec z)\Bigr\}=\exp\Biggl\{-\int_{0}^{t}\lambda_{\rm exc}(x|\vec z)\mathrm{d}x\Biggr\}, 
\label{eq3}
\end{equation}
with $\Lambda_{\rm exc}$ denoting the cumulative excess hazard function.\\
\newpage
\noindent In situations where it is assumed that a fraction of patients will not die from cancer (meaning that $\displaystyle{\lim_{t\rightarrow +\infty}}\Lambda_{\rm exc}(t|\vec z) < +\infty $ or equivalently that $\displaystyle{\lim_{t\rightarrow +\infty}}S_{\rm net}(t|\vec z) > 0$), the observed subjects can be partitioned into two groups (cured and uncured subjects). The net survival can then be expressed as a mixture cure model:
\begin{equation} 
S_{\rm net}(t|\vec z) = \pi(\vec z) S_{1}(t|\vec z) + \{1-\pi(\vec z)\} S_{2}(t|\vec z) = \pi(\vec z) + \{1-\pi(\vec z)\} S_{2}(t|\vec z),
\label{eq4}
\end{equation}
where $S_{1}(t|\vec z) \equiv 1$ and $S_{2}(t|\vec z)$ are the net survival functions of cured  and  uncured patients respectively, the later being a proper survival distribution. The fraction of cured subjects is $\pi(\vec z)$, it depends on covariates $\vec z$ and from~\eqref{eq4}, is equal to $\displaystyle{\lim_{t\rightarrow +\infty}}S_{\rm net}(t|\vec z)$. \\
Although cure models have been originally designed to estimate the fraction of cured subjects, they can be used to estimate another important epidemiological indicator: the time--to--cure i.e. the delay elapsed between the diagnosis and the "time from which cure is reached". Existing cure models do not allow direct estimation of the time--to--cure although this indicator seems crucial; as for instance it can be used to improve the estimation of the delay for the right to be forgotten for cancer survivors. Indeed the right to be forgotten provision is an important milestone in European policymaking. However, it is not universally accessible to cancer survivors across Europe nor does it address all their specific issues. Cancer survivors are often disadvantaged when applying for essential services such as loans, mortgages or child adoption (Youth Cancer Europe, 2018).\\ 
Different methods to estimate the time--to--cure after fitting a cure model have been proposed: Dal Maso et al. (2014) defined the time--to--cure as the delay elapsed between the diagnosis and the time from which the 5--year conditional net survival (defined as the ratio between the net survival at time $t+5$ years and the net survival at time $t$) becomes greater than $0.95$. In a recent paper Boussari et al. (2018) proposed to consider for a given patient $i$, the probability $p_{i}(t)$ of being cured at a given time $t$ after diagnosis knowing that he/she was alive up to $t$ (this probability is nothing but the ratio between the cure fraction and the net survival at time $t$); then the time--to--cure is estimated as the delay from which $p_{i}(t)$ reaches 0.95.\\
This work considers a natural definition of the time--to--cure, named hereafter \textit{time--to--null--excess--hazard} (TNEH), as the delay elapsed between the diagnosis and the time from which the excess hazard becomes null, and proposes a new excess hazard model where the TNEH is a covariate dependent parameter to be estimated. \\
We obtain from~\eqref{eq3},~\eqref{eq4} and the definitions of both the cure fraction and the TNEH: 
 \begin{equation} 
\pi(\vec z) = \displaystyle{\lim_{t\rightarrow +\infty}}S_{\rm net}(t|\vec z) = \exp\Bigl\{-\Lambda_{\rm exc}(\tau(\vec z)|\vec z)\Bigr\} = 
\exp\Biggl\{-\int_{0}^{\tau(\vec z)}\lambda_{\rm exc}(x|\vec z)\mathrm{d}x\Biggr\},
\label{eq5}
\end{equation}
where $\tau(\vec z)$ is the TNEH depending on the covariates $\vec z$ and $\lambda_{\rm exc}(t|\vec z)$=0 whenever $t>\tau(\vec z)$.\\
An illustrative plot of the three hazards functions (described in~\eqref{eq1}) is given in Figure~\ref{fig1} for an individual diagnosed at $55$ or $70$ years, assuming that cure is reached and that TNEH depends on the age at diagnosis. \\
The rest of the paper is organized as follows. In the next Section the new model and its properties are presented as well as the parameters estimation procedure. We illustrate the performances of the estimators derived from our model through both a simulation study in Section~\ref{S3} and applications to survival data from French cancer registries in Section~\ref{S4}. The last Section is devoted to some concluding remarks summarizing the paper and providing some future related researches.
\section{New cure model including the TNEH as parameter} \label{S2} 
\subsection{Model specification}
Various parametric excess hazard functions have been explored to incorporate the TNEH as a parameter ($\tau$). According to the definitions of both the TNEH and the excess hazard function, one of the conditions that must be held by any candidate excess hazard function is to be continuous on $[0, +\infty)$, positive on $[0, \tau)$ and null from $\tau$. Besides, the function must have the ability to reproduce a large panel of  excess hazard curves encountered, in particular in cancer survival study.  The new excess hazard model is written as follows:
\begin{equation} 
\lambda_{\rm exc}(t | \vec z;\vec \theta) = 
\Biggl\{\frac{t}{\tau(\vec z; \vec \eta)} \Biggr\}^{\alpha(\vec z; \vec \gamma)-1} \Biggl\{1 - \frac{t}{\tau(\vec z; \vec \eta)} \Biggr\}^{\beta-1} \mathds{1}_{\{ 0 \leq t \leq \tau(\vec z; \vec \eta) \}},
\label{eq6}
\end{equation}
where
$\tau(\vec z; \vec \eta)>0$ is the TNEH depending on covariates $\vec z$ through the vector of parameters $\vec \eta$. Both $\beta>1$ and $\alpha(\vec z; \vec \gamma)>0$ are shape parameters, the later depending on covariates $\vec z$ through the vector of parameters $\vec \gamma$. Hence 
$\vec \theta=(\vec \gamma, \beta, \vec \eta)$ is the vector of parameters to be estimated.
Note that we constrained $\beta$ to be larger than 1 in order to insure the nullity and the continuity of $\lambda_{\rm exc}(t | \vec z;\vec \theta)$ at $\tau(\vec z; \vec \eta)$.  The shape of the excess hazard function is therefore dependent of the value of $\alpha(\vec z; \vec \gamma)$, either it belongs to $(0,1)$ and the excess hazard function is non increasing on $[0,\tau(\vec z; \vec \eta)]$ with $\lambda_{\rm exc}(t | \vec z;\vec \theta)$ tending to infinity as $t$ tends to 0, or it is larger than 1 and $\lambda_{\rm exc}(t | \vec z;\vec \theta)$ is N-shaped with a maximum located at $\{\alpha(\vec z; \vec \gamma)-1\}/\{\alpha(\vec z; \vec \gamma)+\beta-1\}$; thus because of the linear link between the covariates and the parameters interpreting the effect of covariates in the shape of excess hazard rate seems easy.  Again, the covariates effects on the TNEH ($\tau(\vec z; \vec \eta)$) are easy to interpret because of the linear link. 
In the sequel we refer to the above specified model as the beta--TNEH model.\\
An example of the beta--TNEH model, with the sex (\textit{sex}) and age at the diagnosis of disease (\textit{age}) as covariates can be expressed:
$$ 
\lambda_{\rm exc}(t | \vec z;\vec \theta) = 
\Biggl(\frac{t}{\eta_{0}+\eta_{1} \times age} \Biggr)^{(\gamma_{0}+\gamma_{1} \times age + \gamma_{2} \times sex)-1} 
\Biggl(1 - \frac{t}{\eta_{0}+\eta_{1} \times age} \Biggr)^{\beta-1} 
\mathds{1}_{\{ 0 \leq t \leq (\eta_{0}+\eta_{1} \times age) \}}, 
$$
with $\vec z=(age, sex)$ and $\vec \theta = (\gamma_{0},\gamma_{1},\gamma_{2},\beta,\eta_{0},\eta_{1})$.\\  
As the equation~\eqref{eq6} has the structure of a Beta probability density function, we obtain:\vspace{0.5cm}\\
\textbf{The cumulative excess hazard function} 
\begin{align*} 
\Lambda_{\rm exc}(t|\vec z;\vec \theta)
& = \left \{ \begin{array}{ll}
\tau(\vec z; \vec \eta) \int_{0}^{\frac{t}{\tau(\vec z; \vec \eta)}} 
 x^{\alpha(\vec z; \vec \gamma)-1}(1-x)^{\beta-1} \hspace{0.1cm} \mathrm{d}x &\mbox{ if } 0 \leq t \leq \tau(\vec z; \vec \eta) \nonumber\\  
\tau(\vec z; \vec \eta) \int_{0}^{1} 
 x^{\alpha(\vec z; \vec \gamma)-1}(1-x)^{\beta-1} \hspace{0.1cm} \mathrm{d}x  &\mbox{ if } t > \tau(\vec z; \vec \eta) \end{array} \right.  \nonumber\\
 & = \tau(\vec z; \vec \eta)\hspace{0.1cm} \mathrm{B}\Bigl(\alpha(\vec z; \vec \gamma),\beta\Bigr)\hspace{0.1cm} 
  \mathrm{F}_{\mathcal{B}e}\Bigl(\frac{t}{\tau(\vec z; \vec \eta)};\alpha(\vec z; \vec \gamma),\beta\Bigr),\nonumber
\end{align*}
where $\mathrm{B}$ denotes the beta function and $\mathrm{F}_{\mathcal{B}e}\Bigl( \cdot \hspace{0.1cm} ; \alpha(\vec z; \vec \gamma),\beta\Bigr)$  is the cumulative distribution function of a beta distribution with parameters $\alpha(\vec z; \vec \gamma)$ and $\beta$. \vspace{0.4cm}\\
\textbf{The net survival function} \vspace{-0.2cm}
\begin{align*} 
S_{\rm net}(t|\vec z;\vec \theta) 
 & = \exp \Bigl\{ - \Lambda_{\rm exc}(t|\vec z;\vec \theta) \Bigr\} \nonumber\\
 & = \exp \Biggl\{ - \tau(\vec z; \vec \eta)\hspace{0.1cm} \mathrm{B}\Bigl(\alpha(\vec z; \vec \gamma),\beta\Bigr) 
  \mathrm{F}_{\mathcal{B}e}\Bigl(\frac{t}{\tau(\vec z; \vec \eta)};\alpha(\vec z; \vec \gamma),\beta\Bigr)\Biggr\}. \nonumber 
\end{align*} 

\vspace{0.5cm}\noindent\textbf{The cure fraction} \vspace{-0.3cm}
\begin{align*} 
\pi(\vec z;\vec \theta) 
 & = \exp \Bigl\{ - \Lambda_{\rm exc}\Bigl(\tau(\vec z; \vec \eta)|\vec z;\vec \theta \Bigr) \Bigr\}\nonumber\\
 & = \exp \Bigl\{ - \tau(\vec z; \vec \eta)\hspace{0.1cm} \mathrm{B}\Bigl(\alpha(\vec z; \vec \gamma),\beta\Bigr) \Bigr\},
\nonumber
\end{align*}
hence for covariates that influenced only to the TNEH, it is easy to derive a log-linear link with the cure rate function, otherwise interpretation of the effect of covariates on the cure rate may be more complex.\\

\noindent \textbf{Remark 1}: 
\begin{enumerate}
\item[(i)] From the above two latest results we obtain: 
$ S_{\rm net}(t|\vec z;\vec \theta) = 
\Bigl\{\pi(\vec z;\vec \theta)\Bigr\}^{\mathrm{F}_{\mathcal{B}e}\Bigl(\frac{t}{\tau(\vec z; \vec \eta)};\alpha(\vec z; \vec \gamma),\beta\Bigr)} $. 
We recognize here the form of the NMCM, then the beta--TNEH model can be seen as a special case of the non-mixture cure models family.
\item[(ii)]The issue of covariates incorporation in model~\eqref{eq6} (especially in both the two shape parameters $\alpha$ and $\beta$) is discussed in Section~\ref{S5}. 
\item[(iii)] 
A huge literature about mixture models identifiability exists. Recently general results about the identifiability of parameters of a cure model have been obtained by Hanin and Huang (2014). Because of the specificity of our model we provide in the appendix conditions under which a direct proof of parameters identifiability is obtained.
\end{enumerate}

\subsection{Parameters estimation}
\textbf{Maximum Likelihood Estimator (MLE):} For a subject $i$ we observe $(t_{i},\delta_{i},a_{i},\vec z_{i})$ a realization of $(T_{i},\Delta_{i},A_{i},\vec Z_{i})$. The contribution of the $i^{th}$ subject to the log-likelihood is 
\begin{align} 
\ell(\vec \theta|t_{i},\delta_{i},a_{i},\vec z_{i}) 
 & = \delta_{i}\log\left(\lambda_{\rm obs}(t_{i},a_{i}|\vec z_{i},\vec \theta)\right) - \Lambda_{\rm obs}(t_{i},a_{i}|\vec z_{i},\vec \theta)\nonumber\\
 & \equiv \delta_{i}\log\left(\lambda_{\rm pop}(t_{i}+a_{i}|\vec z_{i}) + \lambda_{\rm exc}(t_{i}|\vec z_{i},\vec \theta)\right) - \Lambda_{\rm exc}(t_{i}|\vec z_{i},\vec \theta). 
\label{eq7}
\end{align}
Hence for a sample of $n$ subjects, the MLE satisfies 
$$
\widehat{\vec \theta}=\arg\max_{\vec \theta\in\Theta}\sum_{i=1}^n\ell(\vec \theta|t_{i},\delta_{i},a_{i},\vec z_{i}).
$$
\textbf{Standard errors of the MLE:} It is well known (see e.g. Newey and McFadden (1994)) that
$\sqrt{n}\left(\widehat{\vec \theta}-\vec \theta_0\right)\stackrel{d}{\longrightarrow} {\cal N}\left(0,I_{0}^{-1}\right)$ as $n$ tends to infinity,
where $I_{0}$ denotes the Fisher information matrix. 
Using standard martingale methods for counting processes (see Andersen et al. (2012), Section VI.1), $I_{0}$ is consistently estimated by $\widehat I$ defined by  
\begin{equation}
\widehat I=\frac{1}{n}\sum_{i=1}^n\left\{\frac{\delta_i \frac{\partial \lambda_{\rm exc}}{\partial \vec \theta}(t_i|\vec z_i , \widehat{\vec \theta})}
{\lambda_{\rm pop}(t_i+a_i|\vec z_i)+\lambda_{\rm exc}(t_i|\vec z_i , \widehat{\vec \theta})}\right\}^{\otimes 2},
\label{eq8}
\end{equation}
where for a column vector $\vec v$, $\vec v^{\otimes 2}=\vec v\vec v^T$. Thus the standard error of the $i$-th component of $\widehat{\vec \theta}$ is estimated by the square root of the $i$-th diagonal entry of $n^{-1}\widehat I^{-1}$. Moreover the standard errors of other quantities related to $\widehat{\vec \theta}$ such as the cure fraction or the net survival could be derived easily using the delta method.\\

\noindent \textbf{Remark 2}: \\
In the estimation  procedure we use the bound constrained optimization method (L-BFGS-B) of Byrd et al. (1995). This method takes the advantage of the BFGS algorithm which is shown to have good performance even for non-smooth optimization functions (Lewis and Overton, 2009), and uses simple bounds constraints. 
The optimization algorithm looks for a $\widehat{\vec\theta}$ belonging to a predefined set having the form $\displaystyle\prod_{j=1}^{k}[\theta_{j,min} \hspace{0.1cm} , \hspace{0.1cm} \theta_{j,max}] \subset \mathbb{R}^{k}$ where $k \in \mathbb{N}$ denotes the number of the parameters. The initial value of $\vec\theta$ must belong to the predefined set. 
A help to set the boundaries is for instance, informations derived from the model constraints ($\tau>0$, $\beta>1$, $\alpha>0$, see~\eqref{eq6}). Note that if any of the estimates equals a boundary, the predefined set of boundaries must be expanded, followed by a new estimation of the parameters. We provide in Web Appendix A, the sets of boundaries used in the estimations steps of the following numerical studies (Sections~\ref{S3} and~\ref{S4}).   
The L-BFGS-B method is already implemented in the R software, package \textit{stats}, function \textit{optim}, (R Core Team, 2019); it requires very short time to run. 

\section{Simulation study \label{S3}}
\subsection{Simulated examples \label{S3.1}}
For the data generation algorithm, one can refer to Web Appendix B.  \\ 
In the following the model complexity was reduced, without loss of generality, by considering only the age at diagnosis as covariate. Hence we consider the following beta--TNEH model:
\begin{equation}
\lambda_{\rm exc}(t | a;\vec \theta) = 
\Biggl\{\frac{t}{\tau(a; \vec \eta)} \Biggr\}^{\alpha(a; \vec \gamma)-1} \Biggl\{1 - \frac{t}{\tau(a; \vec \eta)} \Biggr\}^{\beta-1} \mathds{1}_{\{ 0 \leq t \leq \tau(a; \vec \eta) \}},
\label{eq9}
\end{equation} %
where $\tau(a; \vec \eta) = \eta_{0} + \eta_{1} \times a^{*} $  and $\alpha(a; \vec \gamma) = \gamma_{0} + \gamma_{1} \times a^{*}$ with $a^{*}$ the age at diagnosis standardized using the mean and the standard deviation of its specified distribution. The vector of unknown  parameters, to be estimated from a sample of size $n$,  is  $\vec \theta = (\gamma_{0}, \gamma_{1}, \beta, \eta_{0}, \eta_{1})$.
The population hazard (expected hazard) is assumed to follow a Weibull distribution, the scale and shape parameters being $75$ and $11$ respectively. 
We considered three different settings for the simulations: one illustrating a low excess hazard with a short TNEH and a high censoring rate, the second illustrating a high excess hazard with a moderate TNEH and a low censoring rate, the third illustrating a low excess hazard with a longer TNEH (the excess hazard becomes null very later and the net survival decreases slowly) and a moderate censoring rate. A graphical illustration of the discrepancy between the three settings is provided in Web Appendix C.\\
In the first setting, the vector of true parameters is $\vec \theta = $(2.3, -0.1,  4.8,  5.5,  0.9). The age at diagnosis is uniformly distributed on intervals $[20, 40)$, $[40, 65)$ and $[65, 80]$, and the proportions of age at diagnosis coming from these three intervals are  0.36, 0.29 and 0.35 respectively. The maximum follow-up time (from diagnosis) is fixed to 15 years and the censoring rate is about $60\%$.
\noindent The MLE performances for several sample sizes $n \in \{250, 500, 1000, 2000 \}$ are reported in Table~\ref{Tab1}, Setting 1. 
The bias decreases as the sample size $n$ increases and becomes very small for $n=2000$.
The discrepancies between the standard deviations of the $n$ estimates ($sd$) and the empirical means of the standard deviation estimates ($\overline{se^{*}}$) are low, particularly when $n=2000$. Furthermore, the standard deviations are reduced by half when the sample size $n$ is quadrupled showing that the root--of--$n$ asymptotic convergence rate is reached. For $n=2000$ the coverage probabilities ($cp$) are close to 0.95 which is another indicator that the MLE behaves well for a sample of size $2000$. \\
\noindent Table~\ref{Tab1}, Setting 2, summarizes the MLE performances from the second setting of simulation. Here  the vector of true parameters is $\vec \theta = $(1.25, -0.05, 3.5, 9, 0.3). The age at diagnosis is uniformly distributed on intervals $[20, 50)$, $[50, 70)$ and $[70, 80]$, and the proportions of age at diagnosis coming from these three intervals are  0.15, 0.60 and 0.25 respectively. The maximum follow-up time is fixed to fifteen years and the censoring rate is about $20\%$.
\noindent The MLE performances from the third setting of simulation are summarized in Table~\ref{Tab1}, Setting 3, where the vector of true parameters is $\vec \theta = $(3.01, -0.2, 2.98, 18, 1.2). The age at diagnosis is distributed as in the first setting, the maximum follow-up time is fixed to twenty--five years and the censoring rate is about $46\%$.  
Overall the MLE performances are slightly better in the first and second settings than in the third setting. 
The findings from the first simulation setting are consolidated showing that good estimates can be obtained with the MLE even with moderate sample sizes. Similar results regarding the MLE performances of the beta--TNEH model were obtained when a more extensive simulation setting was considered with $\lambda_{\rm exc}$ depending on three covariates: age at diagnosis (continuous), sex, stage of cancer (3 stages = I, II and III). See Web Appendix D for full details. \\
Moreover, we checked the beta--TNEH model performances when data were generated from distributions that do not fulfill the assumptions underlying the new model. We fit the beta--TNEH model to data simulated from another cure model, the Weibull mixture cure model (a mixture cure model where the survival time of uncured subjects follows a Weibull distribution) and assess performances of the beta--TNEH model by computing the bias, the root mean square error and the coverage probability for the cure fraction and for the net survival at time $t=$ 5, 10 and 15 years. Results were provided in the appendix E. Overall, the beta--TNEH model gives good estimations only if the excess hazard from the Weibull mixture cure model becomes almost null within the follow-up interval.\\
We note that when data are generated from a distribution that do not allow cure fraction (i.e. the cure fraction is null), the beta--TNEH model does not fit the data and gives poor estimations. Indeed, a null cure fraction hypothesis is not compatible with the beta--TNEH model because it corresponds to a TNEH equal to infinity which is outside of the area of validity of the model.
\begin{table}
\caption{MLE performances for various sample sizes based on 1000 simulated samples: mean is the empirical mean, $sd$ is the empirical standard deviation, $\overline{se^{*}}$ is the mean of the  standard errors estimates and \textit{cp} is the 95\% coverage probability. The censoring rates are about 60\% (Setting 1), 20\% (Setting 2) and 46\% (Setting 3).}
\label{Tab1}
\begin{center} 
\renewcommand{\arraystretch}{0.7}
\begin{tabular}{p{2cm} r r r r r r r}
\hline
\hline 
\textit{Setting 1}  & $n$ & $indicators$ & $\gamma_{0}=2.3$ & $\gamma_{1}=-0.1$ & $\beta=4.8$ & $\eta_{0}=5.5$ & $\eta_{1}=0.9$ \\
\cline{2-8}
 & 250 & $mean$ & 2.688 & $-0.122$ & 4.249 & 5.084 & 1.031   \\  
& & $sd$ & 0.651 & 0.325 & 1.941 & 2.406 & 1.264  \\  
& & $\overline{se^{*}}$ & 0.618 & 0.329 & 2.674 & 3.143 & 1.332  \\
& & $cp$ & 0.965 & 0.988 & 0.722 & 0.730 & 0.858   \\
\cline{2-8}
 &500 & $mean$ & 2.457 & $-0.109$ & 4.595 & 5.388 & 0.995   \\  
& & $sd$ & 0.341 & 0.170 & 1.547 & 1.900 & 0.866  \\  
& & $\overline{se^{*}}$ & 0.331 & 0.167 & 1.747 & 2.142 & 0.878  \\
& & $cp$ & 0.959 & 0.978 & 0.818 & 0.808 & 0.901   \\
 \cline{2-8}
& 1000 & $mean$ & 2.375 & $-0.105$ & 4.733 & 5.487 & 0.940  \\
& & $sd$ & 0.214 & 0.110 & 1.168 & 1.488 & 0.629  \\
& & $\overline{se^{*}}$ & 0.211 & 0.110 & 1.141 & 1.400 & 0.600 \\
& & $cp$ & 0.949 & 0.961 & 0.861 & 0.864 & 0.915  \\ 
 \cline{2-8}
& 2000 & $mean$ & 2.333 & $-0.100$ & 4.756 & 5.481 & 0.918  \\
& & $sd$ & 0.136 & 0.070 & 0.723 & 0.907 & 0.411  \\
& & $\overline{se^{*}}$ & 0.140 & 0.070 & 0.748 & 0.907 & 0.400  \\ 
& & $cp$ & 0.964 & 0.956 & 0.920 & 0.910 & 0.935  \\
 \hline
\textit{Setting 2} & $n$ & $indicators$ & $\gamma_{0}=1.25$ & $\gamma_{1}=-0.05$ & $\beta=3.5$ & $\eta_{0}=9$ & $\eta_{1}=0.3$ \\
\cline{2-8} 
& 250 & $mean$ & 1.260 & $-0.052$ & 3.643 & 9.634 & 0.447   \\  
   &  & $sd$ & 0.052 & 0.038 & 1.377 & 3.913 & 1.080  \\  
& & $\overline{se^{*}}$ & 0.055 & 0.039 & 1.715 & 4.928 & 1.201  \\
& & $cp$ & 0.956 & 0.949 & 0.851 & 0.848 & 0.965   \\
 \cline{2-8}
&  500 & $mean$ & 1.258 & $-0.052$ & 3.466 & 9.027 & 0.320   \\  
   &  & $sd$ & 0.035 & 0.025 & 0.831 & 2.308 & 0.664  \\  
& & $\overline{se^{*}}$ & 0.037 & 0.027 & 0.914 & 2.557 & 0.704  \\
& & $cp$ & 0.952 & 0.959 & 0.870 & 0.859 & 0.961   \\
 \cline{2-8}
& 1000& $mean$ & 1.254 & $-0.051$ & 3.517 & 9.110 & 0.331  \\
  &   & $sd$ & 0.026 & 0.019 & 0.601 & 1.716 & 0.468  \\
& & $\overline{se^{*}}$ & 0.026 & 0.019 & 0.620 & 1.730 & 0.474 \\
& & $cp$ & 0.947 & 0.940 & 0.905 & 0.901 & 0.955   \\ 
 \cline{2-8}
&2000 & $mean$ & 1.252 & $-0.050$ & 3.487 & 8.992 & 0.322 \\
  &   & $sd$ & 0.018 & 0.013 & 0.419 & 1.160 & 0.319    \\
& & $\overline{se^{*}}$ & 0.018 & 0.013 & 0.415 & 1.145 & 0.319   \\ 
& & $cp$ & 0.953 & 0.946 & 0.926 & 0.921 & 0.957   \\
 \hline 
\textit{Setting 3}& $n$ & $indicators$ & $\gamma_{0}=3.01$ & $\gamma_{1}=-0.20$ & $\beta=2.98$ & $\eta_{0}=18$ & $\eta_{1}=1.2$ \\
\cline{2-8} 
&  250 & $mean$ & 3.198 & $-0.196$ & 2.843 & 17.350 & 1.125   \\  
   &  & $sd$ & 0.390 & 0.259 & 0.454 & 3.513 & 2.360  \\  
& & $\overline{se^{*}}$ & 0.359 & 0.249 & 0.434 & 3.300 & 2.338  \\
& & $cp$ & 0.952 & 0.952 & 0.855 & 0.829 & 0.909   \\
\cline{2-8} 
&  500 & $mean$ & 3.081 & $-0.202$ & 2.925 & 17.854 & 1.265   \\  
   &  & $sd$ & 0.257 & 0.154 & 0.329 & 2.623 & 1.788  \\  
& & $\overline{se^{*}}$ & 0.228 & 0.158 & 0.301 & 2.341 & 1.659  \\
& & $cp$ & 0.931 & 0.964 & 0.880 & 0.864 & 0.918   \\
 \cline{2-8}
& 1000& $mean$ & 3.046 & $-0.203$ & 2.943 & 17.909 & 1.259  \\
  &   & $sd$ & 0.162 & 0.110 & 0.202 & 1.664 & 1.1490  \\
& & $\overline{se^{*}}$ & 0.155 & 0.108 & 0.204 & 1.602 & 1.138 \\
& & $cp$ & 0.944 & 0.946 & 0.921 & 0.927 & 0.939   \\ 
 \cline{2-8}
&2000 & $mean$ & 3.021 & $-0.205$ & 2.967 & 18.005 & 1.268 \\
  &   & $sd$ & 0.109 & 0.076 & 0.148 & 1.177 & 0.792    \\
& & $\overline{se^{*}}$ & 0.106 & 0.075 & 0.142 & 1.135 & 0.804   \\ 
& & $cp$ & 0.949 & 0.945 & 0.923 & 0.926 & 0.950   \\
 \hline
\end{tabular}
\end{center}
\end{table}
\subsection{Sensitivity to the initial value in the optimization procedure \label{S3.2}}
\begin{figure}
\centerline{\includegraphics[width=15.5cm,height=6.2cm]{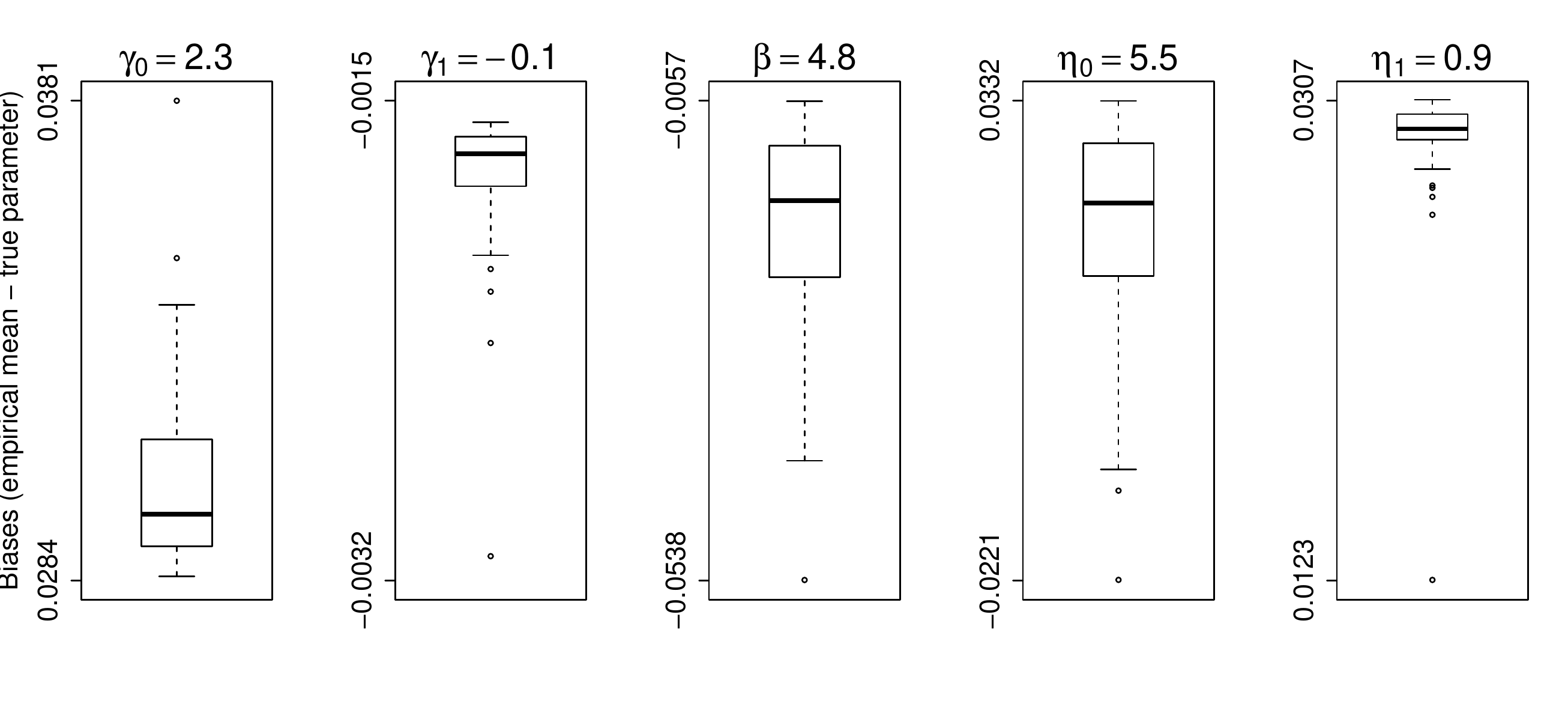}}
\caption{Distributions, based on 1000 simulated samples of size 2000, of the biases  when using $K=30$ initial values in the optimization procedure.}
\label{fig2}
\end{figure}

In this Section we investigate whether the MLE is robust with respect to the chosen initialization point for likelihood maximization algorithm. We then have to verify if the maximization algorithm converges to the same value $\widehat{\vec \theta}$ whatever the chosen initial value. We do this through a simulation study considering again the simulated example 1 with $B=1000$ repetitions of a sample data of size $n=2000$. We generate $K$ initial values of $\vec \theta$ following a multiple uniform distribution on a given space defined by the bounds fixed for the optimization method (see Remark 2). For a given initial value $\vec \theta^{(0)}_{k}$, $k=1,\dots,K$, we compute the empirical mean $\widehat{\vec \theta}_{k}$ of the $B$ estimates of $\vec \theta$, obtained from the $B$ simulated samples data respectively. Then we are interested in the biases between the $\widehat{\vec \theta}_{k}$, $k=1,\dots,K$ and the true parameter.\\
\noindent Figure~\ref{fig2} shows for $K=30$ initial values the boxplots obtained from the K estimates of biases between the empirical means and the true parameter. 
The ranges of the computed biases are 0.009, 0.001, 0.048, 0.055 and 0.018 for $\gamma_{0}=2.3$, $\gamma_{1}=-0.1$, $\beta=4.8$, $\eta_{0}=5.5$ and $\eta_{1}=0.9$ respectively (relative biases vary from 0.4\% to 2\%). These values are very low (near 0) which means that the estimates of $\vec \theta$ are almost identical whatever the chosen initial value. Thus the estimates are robust to the choice of the initial point of the optimization algorithm.
\newpage
\section{Illustrative examples on real data \label{S4}}
Data were provided by the French network of cancer registries (FRANCIM). The analysis included all patients diagnosed from 1995 to 2010, aged 15 to 74 years at diagnosis and followed up to June 30, 2013. The follow-up time was censored at fifteen years. The variable "age at diagnosis" was categorized as in previous analyzes of the FRANCIM data (Cowppli-Bony et al., 2016). Thus we define $\tau(a; \vec \eta) = \eta_{0} + \eta_{1} \times \mathds{1}_{A_1}(a) + \dots + \eta_{J} \times \mathds{1}_{A_J}(a)$  and $\alpha(a; \vec \gamma) = \gamma_{0} + \gamma_{1} \times \mathds{1}_{A_1}(a) + \dots + \gamma_{J} \times \mathds{1}_{A_J}(a)$ where $A_j$ ($0 \leq j \leq J$) are the age at diagnosis groups and, $\eta_{j}$ and $\gamma_{j}$ ($1 \leq j \leq J$) denote respectively the deviations from the effects $\eta_{0}$ and $\gamma_{0}$  of age at diagnosis in the reference group $A_0$. In the examples $A_0$ was the group with the largest size. Model~\eqref{eq9} was fitted on two data sets. Expected mortality rates were derived from the observed mortality rates in the general population available by sex, annual age, year of death, Department of residence and provided by the Institut National de la Statistique et des Etudes Economiques - France (see Web Appendix F for more details and plots of the expected mortality).
\subsection{Testicular cancer data}
\begin{figure}
\centerline{\includegraphics[width=20cm,height=12.5cm]{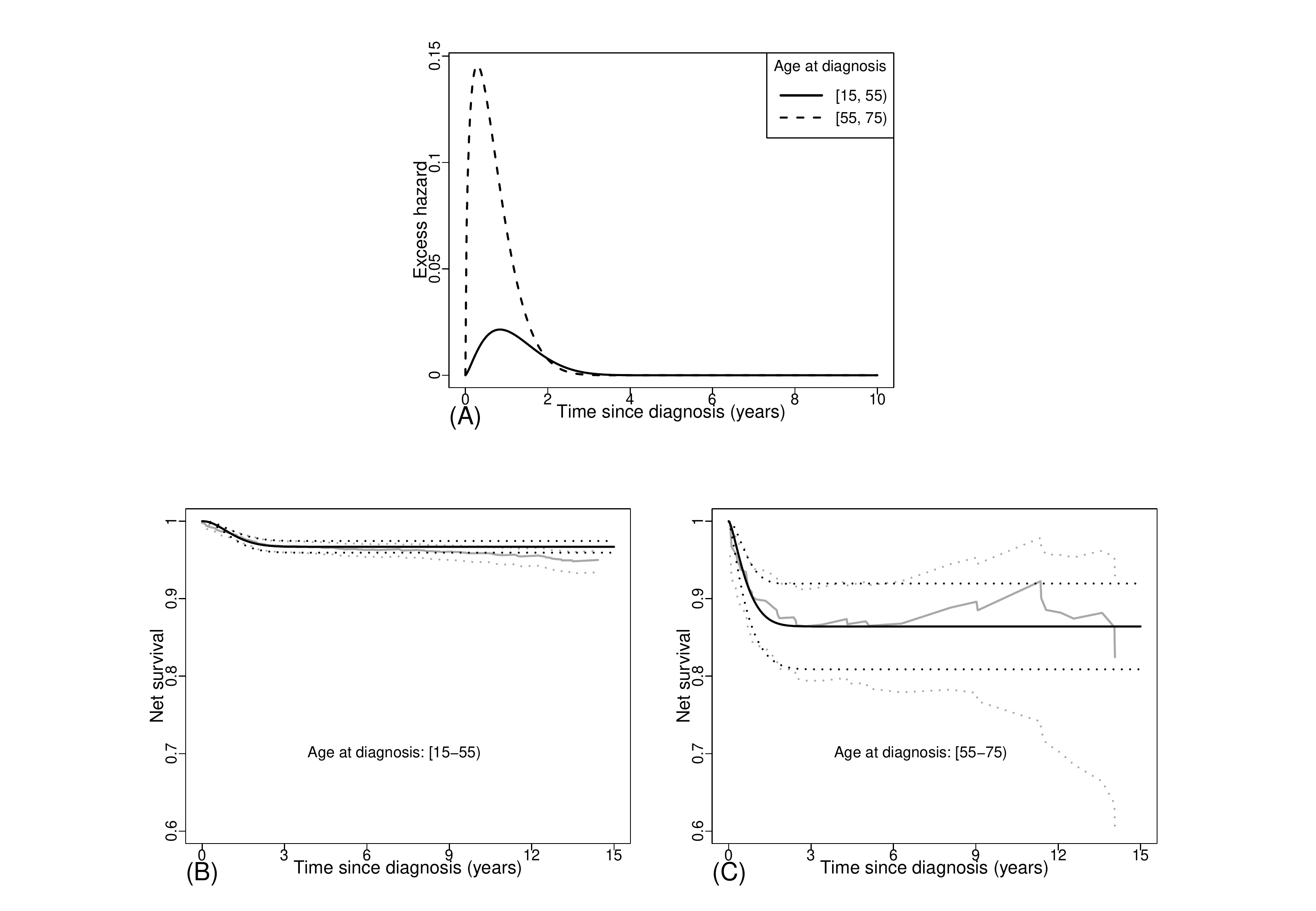}}
\caption{Testicular cancer: estimated Excess hazards function (A) by the beta--TNEH model, and estimated Net survivals functions with their estimated confidence bounds (B and C) from both the PP model (gray lines) and the beta--TNEH model (black lines).}
\label{fig3}
\end{figure}

We considered data of 2834 subjects diagnosed with testicular cancer, for which excess mortality is low. Death was observed for 182 subjects ($6.4\%$ of the cohort) and the observed median survival time since diagnosis for fatal cases was 2.13 years. \\
Age at diagnosis was categorized into 4 groups $A_0=$ [15,45), $A_1=$ [45,55), $A_2=$ [55,65) and $A_3=$ [65,75), and the unknown parameters as well as their standard errors (in brackets) were estimated: $\widehat{\gamma_0}=2.41(0.16)$, $\widehat{\gamma_1}=-0.13(0.18)$, $\widehat{\gamma_2}=-0.90(0.15)$, $\widehat{\gamma_3}=-0.84(0.28)$, $\widehat{\beta}=8.40(2.06)$, $\widehat{\eta_0}=5.47 (1.49)$, $\widehat{\eta_1}=-1.07 (0.98)$, $\widehat{\eta_{2}}=-0.30(0.92)$, $\widehat{\eta_{3}}=-2.94(1.64)$; with a log-likelihood equal to $-1152.12$, corresponding to an Akaike Information Criterion (AIC) equal to $2322.24$. \\
According to the above estimations we pooled $A_0$ and $A_1$ as well as $A_2$ and $A_3$ and fitted another model with age at diagnosis categorized into 2 groups $A_0=$ [15,55) and $A_1=$ [55,75). The estimates were then: $\widehat{\gamma_0}=2.39(0.15)$, $\widehat{\gamma_1}=-0.87(0.14)$, $\widehat{\beta}=8.40(2.09)$, $\widehat{\eta_0}=5.31 (1.45)$  and $\widehat{\eta_1}=-0.91 (0.81)$ and the log-likelihood $= -1153.87$, corresponding to an AIC $= 2317.74$. 
Based on the AIC values we selected the model with 2 age groups. \\
Figure~\ref{fig3}A shows the resulting excess hazard functions while the net survivals from the model with 2 age groups are given in Figure~\ref{fig3} B and C. The excess hazard peaks around the first year after diagnosis whatever the age group. A significant discrepancy is observed between the two excess hazards during the two first years after diagnosis; they become almost identical from two years after diagnosis and close to zero a year later. For each of the age group, we plotted in the same panel (Figure~\ref{fig3} B and C), the net survival (with confidence bounds) estimation by the beta--TNEH model and the nonparametric estimation of the net survival (with confidence bounds) by the method of Perme, Stare, and Est\`eve (2012), denoted hereafter PP model. The two curves match well enough whatever the age group showing that the beta--TNEH model provides a reasonable description of the data. 
Hence from the testicular cancer data, the cure fraction estimates followed by their 95\% confidence intervals were 0.97 [0.93, 1.00] and 0.86 [0.72, 1.00] in the [15, 55) and the [55, 75) age at diagnosis groups respectively. The TNEH estimate in the [15, 55) age at diagnosis group is $\widehat{\eta_0}=5.31$ years with a 95\% confidence interval equal to [2.45, 8.17]. In the [55, 75) age at diagnosis group, the TNEH estimate is $\widehat{\eta_0}+\widehat{\eta_1}=4.40$ years with a 95\% confidence interval equal to [1.91, 6.89]. \\
When fitting a flexible parametric cure model (Andersson et al., 2011) to the testicular cancer data, the derived time--to--cure estimates from Dal Maso et al. (2014) approach with a cut-off at 0.95 were $0$ year and $2.1$ years in the [15, 55) and the [55, 75) age at diagnosis groups respectively.  

\newpage

\subsection{Women pancreatic cancer data}
\begin{figure}
\centerline{\includegraphics[width=16.5cm,height=20.5cm]{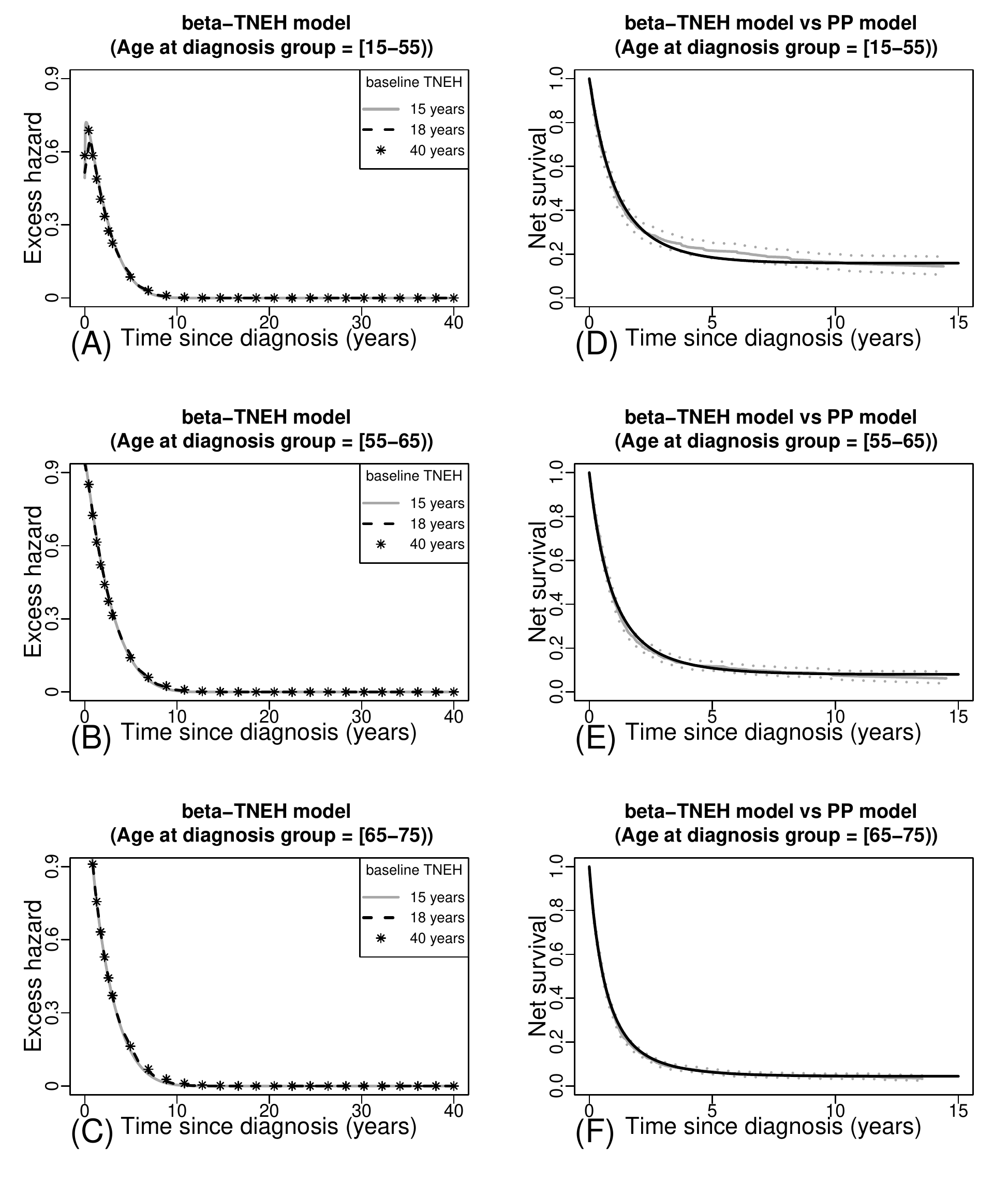}}
\caption{Women pancreatic cancer: on the left, the beta--TNEH model estimations of the excess hazards (panels A, B and C, one panel for each age at diagnosis group) when the TNEH's baseline estimate equals its upper bound fixed to 15 (solid gray line), 18 (dashed black lines) and 40 years (asterisk). On the right, the net survivals estimations for each age at diagnosis group (panels D, E and F, one panel for each age at diagnosis group) by the PP model with confidence bounds (solid gray and dotted gray lines) and by the beta--TNEH model (solid black lines) where the TNEH's baseline estimate equals 40 years.}
\label{fig4}
\end{figure}

We considered data of 3239 women diagnosed with pancreatic cancer, which is a cancer with a very high excess mortality. Death was observed for $91.6\%$ of the subjects and half of the deaths were observed at least 0.6 years after diagnosis. The median and mean age at diagnosis were 66 years and 63.17 years respectively. We fitted a model with the age at diagnosis categorized into 3 groups $A_0=$ [65,75), $A_1=$ [55,65) and $A_2=$ [15,55). \\
In the estimation procedure (see Remark 2), the upper bound $\eta_{0,max}$ for the search of the baseline TNEH (i.e. TNEH for the reference age group $A_0$) was first fixed to 15 years (corresponding to the maximum follow-up time) and we observed that the estimate $\widehat{\eta_0}$ equals the upper bound. When varying $\eta_{0,max}$ up to 40 years, the estimation $\widehat{\eta_0}$ still reached the upper bound. According to the fixed value for $\eta_{0,max}$, the estimations of the other parameters varied but were always different from both their lower and upper bounds fixed for the estimation procedure.
The parameter estimates and standard errors as well as the log-likelihood of the fitted model for three different values of $\eta_{0,max}$ (i.e. $\eta_{0,max}=15, 18, 40$ years) are reported in Table~\ref{Tab2}. The likelihood increased with $\widehat{\eta_{0}}$, and for each parameter the standards errors estimates showed that the derived three confidence intervals overlapped ($95\%$ confidence intervals are $[ 9.73,  20.27]$, $[9.68, 26.32]$ and $[-13.37, 93.37]$ respectively). Note that for $\widehat{\eta_0} = 40$ years, we are far out of the follow-up interval (0 to 15 years) meaning that no data are available for consistent estimations of the parameters; this leads to large standard errors and consequently possible negative value for the lower bound of the $95\%$ confidence interval, what must be interpreted with caution. For each age at diagnosis group, the  3 excess hazards corresponding to $\widehat{\eta_0}=15, 18$ and $40$ years are plotted in Figure~\ref{fig4}, panels A, B and C. Whatever the age group, the 3 curves were very similar showing that the changes in the parameter estimates had little impact on the corresponding excess hazards. The panels D, E and F of Figure~\ref{fig4} show for the 3 age groups, both the parametric and the nonparametric estimations of the survival function using the parametric beta--TNEH model and the nonparametric PP model. Overall whatever the age group, the two curves matched well, meaning that the beta--TNEH model provided a reliable estimation of the net survival function even if the TNEH was obviously underestimated.
Note that the beta--TNEH model was able to provide satisfactory estimations because of the behavior of the excess hazard function which was high just after diagnosis, decreased rapidly and became very close to zero before the end of the follow-up.   
Although the net survival shows an apparent plateau (often translated as the existence of cure), this example shows a situation where either the TNEH had a non-finite value or the TNEH was too large, greater than a predefined threshold (for instance the maximum follow-up time) over which the estimated TNEH has no practical usefulness.\\
On the other hand, the approaches by both Dal Maso et al. (2014) and Boussari et al.(2018) provided approximations of the time--to--cure for the women pancreatic cancer data after fitting a flexible parametric cure model (Andersson et al., 2011). The time--to--cure estimates from the approach of Dal Maso et al. (2014) using a cut-off at 0.95, were 9 years in both the [15, 55) and the [55, 65) age at diagnosis groups and 10 years in the [65, 75) age at diagnosis group; very close results were obtained from the approach proposed by Boussari et al. (2018) (see the related paper for more details).
\begin{table}
\caption{Parameter estimates with standard errors in brackets and log-likelihood (LL) when  the upper bound of $\eta_0$ (the baseline TNEH) is fixed to 15, 18 and 40 years.}
\label{Tab2}
\begin{center}
\begin{tabular}{c c c c c c c c c}
\hline
\hline
$\eta_{0,max}$ & LL & $\widehat{\gamma_{0}}$ & $\widehat{\gamma_{1}}$ & $\widehat{\gamma_{2}}$ & $\widehat{\beta}$ & $\widehat{\eta_{0}}$ & $\widehat{\eta_{1}}$  & $\widehat{\eta_{2}}$\\
\hline
15 & $-3741.758$ & 0.931 & 0.074 & 0.128 & 6.005 & 15.000 & 0.259 & $-1.959$\\  
&   & (0.007) & (0.013) & (0.018) & (1.093) & (2.690) & (0.962) & (1.039)   \\
\hline     
18 & $-3737.123$ & 0.933 & 0.071 & 0.121 & 7.156 & 18.000 & 0.282 & $-2.608$ \\  
 &  & (0.007) & (0.013) & (0.018) & (1.712) & (4.243) & (1.208) & (1.390)   \\
\hline  
40 & $-3727.847$ & 0.941 & 0.058 & 0.100 & 15.650 & 40.000 & $-0.535$ & $-6.850$ \\  
&  & (0.008) & (0.013) & (0.021) & (10.848) & (27.234) & (2.930) & (5.523) \\ 
 \hline
\end{tabular}
\end{center}
\end{table}

\newpage
\section{Concluding remarks \label{S5}}
The time from which no more death occurs from a disease of interest (such as cancer) is a useful indicator in epidemiological studies, and can help to improve access to insurance and loans for people living with a personal history of cancer. In this paper we refer to this delay as the time--to--null--excess--hazard (TNEH). While sophisticated models have been proposed to estimate efficiently the fraction of cured patients, it seems that there is a lack of methods in the literature for the TNEH estimation. \\
We proposed a cure model based on a paradigm where the excess hazard function includes the TNEH as a parameter to be estimated. The proposed beta--TNEH model could be treated as a special case of the NMCM. The simulation study showed that the beta--TNEH presented good performances regarding the maximum likelihood estimation method. However we advise to do not use the new method when cure assumption is not reasonable because this will lead to poor estimations with meaningless results; the beta--TNEH model is really suitable for data showing an excess hazard which becomes null within the follow-up interval. Existing methods (for instance Dal Maso et al., 2014 and Boussari et al., 2018) can provide an approximation of the time--to--cure when the excess hazard becomes just relatively low and not necessarily null. But these methods are based on approximations requiring a choice of a cut-off what could easily be subject of criticism since the derived time--to--cure estimate could be very sensitive to the predefined value of the cut-off.\\ 
Two examples on real data sets were treated and for each of them, the net survival estimated by the beta--TNEH model was very close to the estimation provided by the nonparametric PP model. With testicular cancer data, robust finite value of the TNEH was estimated. With women pancreatic cancer data, despite the fact that the excess hazard becomes almost zero around 10 years after diagnosis, the TNEH's baseline estimate equals the corresponding upper bound specified for the optimization procedure, even for a relatively large value of the upper bound (40 years). In such situations, one can conclude at least that the TNEH is greater than a predefined time $\mathrm{T^{*}}$ having practical usefulness. Hence treating the TNEH as a parameter would offer a way to test a cure hypothesis. Of course, by testing the hypotheses "TNEH $\leq \mathrm{T^{*}}$" versus "TNEH $> \mathrm{T^{*}}$", one could conclude at least whether there is evidence of cure before $\mathrm{T^{*}}$ or not. \\
Due to its similarity with a Beta probability distribution function, the new model could reproduce various excess hazard curves, offers a simple and comprehensible way to handle the TNEH as a model parameter and leads to satisfactory estimates as shown by the numerical studies. However, in some cases, it would not be flexible enough to capture some shapes for instance in situation where the excess hazard shows several local extrema before reaching zero. Thus the new paradigm should be adapted to more flexible cure models to account for a larger panel of hazard curves. For instance in the "flexible" NMCM model proposed by Andersson et al. (2011), the baseline cumulative excess hazard is modeled by a restricted cubic spline of the time since diagnosis. Instead of fixing arbitrary the last knot of the spline after the last observed event time as they did, one could consider this knot as a parameter corresponding to the TNEH. However, treating a spline knot as a model parameter to be estimated is a topic that would need more research.\\
For model~\eqref{eq6} specification purpose we had considered the case where both the two shape parameters $\alpha$ and $\beta$ were linked to covariates as well as the case where only $\beta$ was linked to covariates. Simulations led to unsatisfactory parameter estimates for these cases (especially when both $\alpha$ and $\beta$ were linked to the same continuous covariates), probably due to numerical problems and/or identifiability issue. This kind of problem seemed inherent to cure models. Indeed, Li, Taylor and Sy (2001) stated that even if a cure model is formally identifiable, a possible near non-identifiability situation could occur as a flat or irregular likelihood surface for finite samples, with associated numerical problems.  Farewell (1982) noted in parametric cure model a high correlation between the parameter estimates of the incidence part of the model and those of the latency part of the model, what we thought, could also lead to numerical problems. More generally, Hanin and Huang (2014) pointed that sharing of covariates between various components of cure models may prevent identifiability and then good parameter estimates.\\
Finally we would like to point out that in the beta--TNEH model, the TNEH is assumed to be deterministic, which means that the model estimates the same value of TNEH for subjects sharing the same characteristics (covariates). A way to improve our model would be to consider the TNEH as a random effect with a specified common distribution depending on covariates. This is an ongoing work. \\  



\section*{Acknowledgements}
The authors like to thank the French network of cancer registries (FRANCIM) for providing the two real data sets of cancers. This work was supported by the Institut National du Cancer (INCa) [grant number 2014--087], by the Fondation ARC pour la recherche sur le cancer [personal grant for author O.B. number PDF20151203665] and by the French Government (Agence Nationale de la Recherche, "Investissements d'Avenir", ANR--11-LABX--0021).\vspace*{-8pt}

\section*{Supporting Information}
Web Appendix A, referenced in Section~\ref{S2}, Web Appendix B, C, D and E
referenced in Section~\ref{S3} and Web Appendix F, referenced in Section~\ref{S4} are available online with this paper.\vspace*{-8pt} \\


\section*{References}
\begin{itemize}
\item[] Andersen, P. K., Borgan O., Gill, R. D., and Keiding, N. (2012). \textit{Statistical Models Based on Counting Processes}. Springer Science \& Business Media.
\item[] Andersen, P. K., Borgan O., Gill, R. D., and Keiding, N. (2012). \textit{Statistical Models Based on Counting Processes}. Springer Science \& Business Media.
\item[] Andersson, T. M. L., Dickman, P. W., Eloranta, S., and Lambert, P. C. (2011). Estimating and modelling cure in population-based cancer studies within the framework of flexible parametric survival models. \textit{BMC Med Res Methodol} 11, 96.
\item[] Boag, J. W. (1949). Maximum likelihood estimates of the proportion of patients cured by cancer therapy. \textit{J R Stat Soc Series B} 11, 15--53.
\item[] Boussari, O., Romain, G., Remontet, L., Bossard, N., Mounier, M., Bouvier, A.-M., Binquet, C., Colonna, M., and Jooste, V. (2018). A new approach to estimate time--to--cure from cancer registries data. \textit{Cancer epidemiol} 53, 72--80.
\item[] Byrd, R. H., Lu, P., Nocedal, J., and Zhu, C. (1995). A limited memory algorithm for bound constrained optimization. \textit{SIAM J Sci Comput} 16, 1190--1208.
\item[] Chen, M.-H., Ibrahim, J. G., and Sinha, D. (1999). A new bayesian model for survival data with a surviving fraction. \textit{J Am Stat Assoc} 94, 909--919.
\item[] Cooner, F., Banerjee, S., Carlin, B. P., and Sinha, D. (2007). Flexible cure rate modeling under latent activation schemes. \textit{J Am Stat Assoc} 102, 560--572.
\item[] Cowppli-Bony, A., Uhry, Z., Remontet, L., Guizard, A.-V., Voirin, N., Monnereau, A., et al. (2016). \textit{Survie des Personnes Atteintes de Cancer en France M\'etropolitaine, 1989--2013. partie 1 -- Tumeurs Solides}. Saint-maurice: Institut de veille sanitaire.
\item[] Dal Maso, L., Guzzinati, S., Buzzoni, C., Capocaccia, R., Serraino, D., Caldarella, A., et al. (2014). Long-term survival, prevalence, and cure of cancer: a population-based estimation for 818,902 italian patients and 26 cancer types. \textit{Ann Oncol} 25, 2251--2260.
\item[] Farewell, V. T. (1982). The use of a mixture model for the analysis of
survival data with long-term survivors. \textit{Biometrics} 38, 1041–1046.
\item[] Gu, Y., Sinha, D., and Banerjee S. (2011). Analysis of cure rate survival data under proportional odds model. \textit{Lifetime Data Anal} 17, 123--134.
\item[] Hanin, L. and Huang, L. S. (2014). Identifiability of cure models revisited. \textit{J Multivar Anal} 130, 261--274.
\item[] Klein, J. P., Van Houwelingen, H. C., Ibrahim, J. G., and Scheike T. H. (Eds.). (2016). \textit{Handbook of Survival Analysis}. CRC Press.
\item[] Kuk, A. Y. C., and Chen, C.-H. (1992). A mixture model combining logistic regression with proportional hazards regression. \textit{Biometrika} 79, 531--541.
\item[] Lambert, P. C., Thompson, J. R., Weston, C. L., and Dickman P. W. (2006). Estimating and modeling the cure fraction in population-based cancer survival analysis. \textit{Biostatistics} 8, 576--594.
\item[] Lehmann, E. L., and Casella, G. (1998). \textit{Theory of point estimation}. 2nd
Edition. New York: Springer-Verlag. 
\item[] Lewis, A. S., and Overton, M. L. (2009). Nonsmooth optimization via bfgs. URL \url{https://cs.nyu.edu/overton/papers/pdffiles/bfgs inexactLS.pdf}.
\item[] Li, C.-S., and Taylor. J. M. G. (2002). Smoothing covariate effects in cure models. \textit{Commun Stat Theory Methods} 31, 477--493.
\item[] Maller, R. A., and Zhou, X. (1996). \textit{Survival Analysis with Long Term Survivors}. John Wiley and sons.
\item[] Newey, W. K., and McFadden, D. (1994). Large sample estimation and hypothesis testing. \textit{Handbook of econometrics} 4, 2111--2245.
\item[] Peng, Y. and Xu, J. (2012). An extended cure model and model selection. \textit{Lifetime Data Anal} 18, 215--233.
\item[] Perme, M. P., Stare, J., and Est\`eve, J. (2012). On estimation in relative survival. \textit{Biometrics} 68, 113--120.
\item[] R Core Team (2019). R: A language and environment for statistical computing. R Foundation for Statistical Computing, Vienna, Austria. URL \url{https://www.R-project.org/}.
\item[] Taylor, J. M. G., and Liu, N. (2007). Statistical issues involved with extending standard models. In \textit{Advances In Statistical Modeling And Inference: Essays in Honor of Kjell A. Doksum} 
299--311. World Scientific.
\item[] Tsodikov, A. D., Ibrahim, J. G., and Yakovlev. A. Y. (2003). Estimating cure rates from survival data: an alternative to two-component mixture models. \textit{J Am Stat Assoc} 98, 1063--1078.
\item[] Verdecchia, A., De Angelis, R., Capocaccia, R., Sant, M., Micheli, A., Gatta, G., and  Berrino F. (1998). The cure for colon cancer: results from the eurocare study. \textit{Int J Cancer} 77, 322--329.
\item[] Yakovlev, A. Y., Tsodikov, A. D., and Asselain, B. (1996). \textit{Stochastic models of tumor latency and their biostatistical applications} (Vol. 1). World Scientific.
\item[] Yin G., and Ibrahim, J. G. (2005a). Cure rate models: a unified approach. \textit{Can J Stat} 33, 559--570.
\item[] Yin G., and Ibrahim, J. G. (2005b). A general class of bayesian survival models with zero and nonzero cure fractions. \textit{Biometrics} 61, 403--412.
\item[] Youth Cancer Europe (2018). The right to be Forgotten for Cancer Survivors. URL \url{https://www.youthcancereurope.org/youth-cancer-news-events/2018/08/event-the-right-to-be-forgotten-for-cancer-survivors/} [accessed 24 August 2019]
\item[] Yu, B., Tiwari, R. C., Cronin, K. A., McDonald, C., and Feuer, E. J. (2005). Cansurv: a windows program for population-based cancer survival analysis. \textit{Comput Methods Programs Biomed} 80, 195--203.
\item[] Zeng, D., Yin, G., and Ibrahim, J.G. (2006). Semiparametric transformation models for survival data with a cure fraction. \textit{J Am Stat Assoc} 101, 670--684.
\item[] Zhang, J., and Peng, Y. (2009). Accelerated hazards mixture cure model. \textit{Lifetime Data anal} 15, 455--467.



\end{itemize}

\appendix


\newpage

\section*{Appendix: Model identifiability \label{A1}}

Let us note $P_{\vec \theta}$ the probability distribution of an observation $(T,\Delta,Z)$ under the parametric model~\eqref{eq6} where for simplicity we consider that the age at diagnosis $A$ is a component of the covariate vector $Z$. According to Lehmann and Casella (1998) the model identifiability holds if $\vec\theta\mapsto P_{\vec \theta}$ is one--to--one on the parameter space $\Theta$. Let us denote by $\mathscr{Z}$ the set of values taken by the covariates. It is straightforward to verify that identifiability of model~\eqref{eq6} is equivalent to the identifiability of the class of functions
$$
\mathscr{H}=\Bigl\{ (t,\vec z)\mapsto \lambda_{\rm exc}( t | \vec z;\vec \theta): [0,+\infty)\times \mathscr{Z}\to [0,+\infty) ; \vec \theta=(\vec \gamma, \beta, \vec \eta) \in \Theta_{\alpha}\times [0,+\infty)\times\Theta_{\tau}\equiv \Theta \Bigr \}
$$
that is $\lambda_{\rm exc}( t | \vec z;\vec \theta)=\lambda_{\rm exc}( t | \vec z;\vec \theta^*)$ for all $(t,\vec z)$ in $[0,+\infty)\times \mathscr{Z}$ implies $\vec\theta=\vec\theta^*$ for all $(\vec\theta,\vec\theta^*)\in\Theta^2$. Let us introduce two additional classes of functions:
\begin{itemize}
\item $\mathscr{A}=\{ \vec z\mapsto \alpha(\vec z;\vec\gamma): \mathscr{Z}\to (0,+\infty); \vec \gamma \in \Theta_{\alpha} \}$;
\item $\mathscr{T}=\{ \vec z\mapsto \tau(\vec z;\vec\eta): \mathscr{Z}\to (0,+\infty); \vec \eta\in \Theta_{\tau} \}$.
\end{itemize}
\noindent \textbf{Proposition 1 } \textit{If the classes of functions $\mathscr{A}$  and $\mathscr{T}$ are identifiable, then model~\eqref{eq6} is identifiable.} \vspace{0.1cm}\\
\noindent\textbf{Proof. } We ever explained that it is sufficient to verify that $\mathscr{H}$ is identifiable. Let us consider $\vec \theta=(\vec \gamma, \beta, \vec \eta) \in \Theta$ and $\vec \theta^*=(\vec \gamma^*, \beta^*, \vec \eta^*) \in \Theta$ such that $\lambda_{\rm exc}( t | \vec z;\vec \theta)=\lambda_{\rm exc}( t | \vec z;\vec \theta^*)$ for all $(t,\vec z)$ in $[0,+\infty)\times \mathscr{Z}$. The supports of $\lambda_{\rm exc}( \cdot | \vec z ;\vec \theta)$ and $\lambda_{\rm exc}( \cdot  | \vec z ;\vec \theta^*)$ having to match for all $\vec z\in \mathscr{Z}$ we have $\tau(\vec z;\vec\eta)=\tau(\vec z;\vec\eta^*)$ for all $\vec z\in \mathscr{Z}$, thus $\vec\eta=\vec\eta^*$ since $\mathscr{T}$ is identifiable. As a consequence, for all $\vec z\in \mathscr{Z}$ and  $t \in (0,\tau(\vec z;\vec\eta))$ we have
$$
\left\{\frac{t}{\tau(\vec z;\vec\eta)}\right\}^{\alpha(\vec z;\vec\gamma)-1}\times \left\{1-\frac{t}{\tau(\vec z;\vec\eta)}\right\}^{\beta-1}=\left\{\frac{t}{\tau(\vec z;\vec\eta)}\right\}^{\alpha(\vec z;\vec\gamma^*)-1}\times \left\{1-\frac{t}{\tau(\vec z;\vec\eta)}\right\}^{\beta^*-1}.
$$
Taking the logarithm of the above equality and using the linear independence of functions $t\mapsto \log\left\{\frac{t}{\tau(\vec z;\vec\eta)}\right\}$ and $t\mapsto \log\left\{1-\frac{t}{\tau(\vec z;\vec\eta)}\right\}$ we obtain $\beta=\beta^*$ and $\alpha(\vec z;\vec\gamma)=\alpha(\vec z;\vec\gamma^*)$ for all $\vec z\in\mathscr{Z}$. The later result with the identifiability of the class of functions $\mathscr{A}$ leads to $\vec\gamma=\vec\gamma^*$. This finishes the proof. \hfill $\Box$

It is easy to check that classes of functions $\mathscr{A}$  and $\mathscr{T}$ that we used for both the simulation study in Section~\ref{S3} and the illustrations on real dataset in Section~\ref{S4} are identifiable.

\label{lastpage}

\end{document}


\textbf{\Large{\centerline{Supporting Information for}}}\\ 
\centerline{Modeling excess hazard with time--to--cure as a parameter} \\
\centerline{by Boussari et al. } 

\newpage
\noindent\textbf{Web Appendix A: boundaries specified in the estimation procedure when using the L-BFGS-B method} \\ 

\noindent $\vec \theta_{up} $ denotes the vector of upper bounds,  $\vec \theta_{lo} $ the vector of lower bounds, $\vec \theta$ the vector of true parameters and $\widehat{\vec \theta}$ the vector of estimated parameters. \\

 \centerline{\textbf{Simulations}} 
\begin{table}[!h]
\begin{center} 
\begin{tabular}{p{2cm} l r r r r r}
\hline
\hline 
\textit{Setting 1}   & $\vec \theta$ & $\gamma_{0}=2.3$ & $\gamma_{1}=-0.1$ & $\beta=4.8$ & $\eta_{0}=5.5$ & $\eta_{1}=0.9$ \\
 & $\vec \theta_{lo}$ & 0.4 & $-3$ & 1 & 1 & $-5$   \\  
 & $\vec \theta_{up} $ & 8 & 3 & 12 & 20 & 5  \\  
\hline 
\textit{Setting 2}   & $\vec \theta$ & $\gamma_{0}=1.25$ & $\gamma_{1}=-0.05$ & $\beta=3.5$ & $\eta_{0}=9$ & $\eta_{1}=0.3$ \\
  & $\vec \theta_{lo}$ & 0.4 & $-3$ & 1 & 1 & $-5$   \\  
 & $\vec \theta_{up} $ & 8 & 3 & 12 & 20 & 5  \\  
\hline 
\end{tabular}
\end{center}
\end{table}
\centerline{\textbf{Testicular cancer}} 
\begin{table}[!h]
\begin{center} 
\begin{tabular}{ l r r r r r}
\hline
\hline 
 $\widehat{\vec \theta}$ & $\widehat{\gamma}_{0}=2.39$ & $\widehat{\gamma}_{1}=-0.87$ & $\widehat{\beta}=8.4$ & $\widehat{\eta}_{0}=5.31$ & $\widehat{\eta}_{1}=-0.91$ \\
$\vec \theta_{lo}$ & 0.4 & $-3$ & 1 & 0.1 & $-5$   \\  
$\vec \theta_{up} $ & 6.4 & 3 & 10 & 15 & 5  \\  
\hline 
\end{tabular}
\end{center}
\end{table}
\centerline{\textbf{Women pancreatic cancer}} 
\begin{table}[!h]
\begin{center}
\begin{tabular}{ l r r r r r r r}
\hline
\hline
$\widehat{\vec \theta}$ & $\widehat{\gamma}_{0}=0.931$ & $\widehat{\gamma}_{1}=0.074$ & $\widehat{\gamma}_{2}=0.128$ & $\widehat{\beta}=6.005$ & $\widehat{\eta}_{0}=15$ & $\widehat{\eta}_{1}=0.259$  & $\widehat{\eta}_{2}=-1.959$\\
$\vec \theta_{lo} $ & 0.4 & $-5$ & $-5$ & 1 & 0.1 & $-10$ & $-10$ \\  
$\vec \theta_{up} $ & 10 & 5 & 5 & 40 & 15 & 10 & 10  \\ 
\hline
$\widehat{\vec \theta}$ & $\widehat{\gamma}_{0}=0.933$ & $\widehat{\gamma}_{1}=0.071$ & $\widehat{\gamma}_{2}=0.121$ & $\widehat{\beta}=7.156$ & $\widehat{\eta}_{0}=18$ & $\widehat{\eta}_{1}=0.282$  & $\widehat{\eta}_{2}=-2.608$\\     
$\vec \theta_{lo} $ & 0.4 & $-5$ & $-5$ & 1 & 0.1 & $-10$ & $-10$ \\  
$\vec \theta_{up} $ & 10 & 5 & 5 & 40 & 18 & 10 & 10  \\
\hline
$\widehat{\vec \theta}$ & $\widehat{\gamma}_{0}=0.941$ & $\widehat{\gamma}_{1}=0.058$ & $\widehat{\gamma}_{2}=0.100$ & $\widehat{\beta}=15.650$ & $\widehat{\eta}_{0}=40$ & $\widehat{\eta}_{1}=-0.535$  & $\widehat{\eta}_{2}=-6.850$\\  
$\vec \theta_{lo} $ & 0.4 & $-5$ & $-5$ & 1 & 0.1 & $-10$ & $-10$ \\  
$\vec \theta_{up} $ & 10 & 5 & 5 & 40 & 40 & 10 & 10  \\ 
 \hline \\
\end{tabular}
\end{center}
\end{table}

\clearpage

\noindent\textbf{Web Appendix B: data simulation algorithm } \\
In the following $\mathcal{U}[a,b]$  denotes the uniform distribution on $[a,b]$ and $\mathcal{B}(p)$ denotes the Bernoulli distribution with parameter $p$. 
\begin{itemize}
\item[(a)] Generate $n$ ($n \in \mathbb{N}$) realizations of age at diagnosis $A$ $(a_{1}, \dots, a_{n})$ and covariates  $\vec Z$ $(\vec z_{1}, \dots, \vec z_{n})$. 
\item[(b)] Generate $n$ realizations of the lifetime $X_{1}$ $(x_{11}, \dots, x_{1n})$ elapsed between the diagnosis and the death due to cancer: draw $n$ realizations of $U$ $(u_{1}, \dots, u_{n})$ with $U \sim \mathcal{U}[0, 1]$, and set  
$x_{1i}=\left \{ \begin{array}{ll}
\tau(\vec z_{i}; \vec \eta)\mathrm{F}_{\mathcal{B}e}^{-1}\Bigl(\frac{\log (u_{i})}{\log \{\pi(\vec z_{i} ; \vec \theta)\}};\alpha(\vec z_{i}; \vec \gamma),\beta\Bigr) &\mbox{ if }u_{i} \geq \pi(\vec z_{i} ; \vec \theta) \\ 
\infty  &\mbox{ if } u_{i} < \pi(\vec z_{i} ; \vec \theta) \end{array} \right.$, for $1\leq i\leq n$.  
\item[(c)] Generate $n$ realizations of the lifetime $X_{2}$ $(x_{21}, \dots, x_{2n})$ elapsed between the diagnosis and the death due to other causes: draw $n$ realizations of $U$ $(u_{1}^{*}, \dots, u_{n}^{*})$ with $U \sim \mathcal{U}[0, 1]$, and set
$x_{2i}=\Lambda_{pop}^{-1}\Bigl[ \log \Bigl\{ \frac{u_i^{*}}{S_{\rm pop}(a_i)}\Bigr\}\Bigr] - a_{i}$, for $1\leq i\leq n$.
\item[(d)] Generate $n$ realizations of $C_{1}$ $(c_{11}, \dots, c_{1n})$ and $n$ realizations of $C_{2}$ $(c_{21}, \dots, c_{2n})$ such that $C_{j}= P_{j} \times U_{j}+(1-P_{j}) \times t_{max}$ for $j \in \{1, 2\}$ where $t_{max}$ denotes a predefined maximum follow-up time, $P_{j} \sim \mathcal{B}(p_{j})$ and $U_{j} \sim \mathcal{U}[0, t_{max}]$. \\
Note that $C_{1}$ describes censoring times corresponding to a mixture of a proportion $p_{1}$ of patients who are still alive at the endpoint of the follow-up and are followed less than $t_{max}$, and a proportion $1-p_{1}$ of patients followed more than $t_{max}$ before the endpoint of the follow-up. 
$C_{2}$ describes censoring times corresponding to a mixture of a proportion $p_{2}$ of patients who are lost to follow-up, and a proportion $1-p_{2}$ of patients followed more than $t_{max}$ before the endpoint of the follow-up. Considering that the censoring time $C$ results from the competing times $C_1$ and $C_2$, the realizations $(c_1,\dots, c_n)$ of the censoring time $C$ are then equal to $c_i=\min(c_{1i},c_{2i})$ for $1\leq i\leq n$.  
\item[(e)] Finally the realizations of the survival time $T$ $(t_1,\dots, t_n)$  and the censoring indicator $\Delta$ $(\delta_1,\dots, \delta_n)$ are obtained by setting $t_i=\min(x_{1i},x_{2i},c_i)$ and $\delta_i=\mathds{1}_{\{ \min(x_{1i},x_{2i})<c_i\}}$, for $1\leq i\leq n$.
\end{itemize} 

\newpage

\noindent\textbf{Web Appendix C: graphical illustration of the simulation settings}\\
 
\begin{figure}[!ht]
\centerline{\includegraphics[width=18cm,height=16cm]{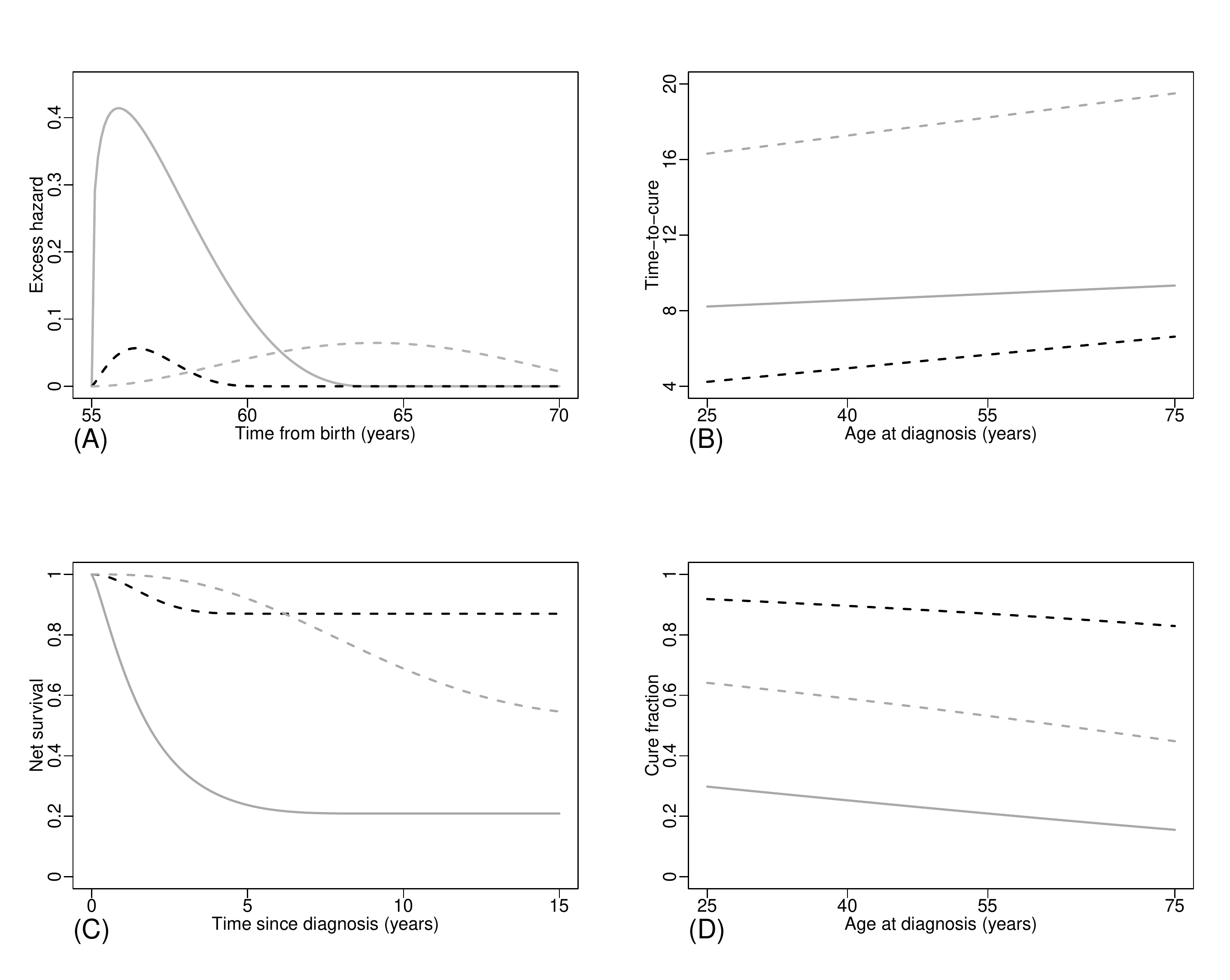}}
\caption*{Excess hazards (A) and Net survivals (C) functions for a patient diagnosed at 55 years; Time--to--null--excess--hazard (B) and Cure fraction (D) according to age at diagnosis. The vector of true parameters are $\vec \theta = (2.3, -0.1,  4.8,  5.5,  0.9)$, $\vec \theta = (1.25, -0.05, 3.5, 9, 0.3)$ and $\vec \theta = (3.01, -0.2, 2.98, 18, 1.2)$ corresponding respectively to the simulation setting 1 (dashed black lines), the simulation setting 2 (grey lines) and the simulation setting 3 (dashed grey lines).}
\end{figure} 

\newpage

\noindent\textbf{Web Appendix D: Performances of a multivariates beta--TNEH model}\\
\noindent We consider the beta--TNEH model (equation (6) in the paper) where $\lambda_{\rm exc}$ depending on three covariates: age at diagnosis (continuous), sex, stage of cancer (3 stages = I, II and III) with $\alpha$ depending on age and stage and $\tau$ depending on age and sex:
$\alpha(\vec z_{1}; \vec \gamma) = \gamma_{0} + \gamma_{1} \times a^{*} + 
\gamma_{2} \times \mathds{1}_{\{\text{stage = II}\}}+ 
\gamma_{3} \times \mathds{1}_{\{\text{stage = III}\}}$ 
and
$\tau(\vec z_{2}; \vec \eta) = \eta_{0} + \eta_{1} \times a^{*} +
\eta_{2} \times \mathds{1}_{\{\text{sex = man}\}}$; 
where $a^{*}$ is the age at diagnosis standardized using the mean and the standard deviation of its specified distribution, $\vec z_{1}=$ (age, stage), $\vec z_{2}=$ (age, sex), 
$\vec \gamma = (\gamma_{0}, \gamma_{1}, \gamma_{2}, \gamma_{3})$ and 
$\vec \eta = (\eta_{0}, \eta_{1}, \eta_{2})$. 
Hence  $\vec \theta = (\vec \gamma, \vec \eta)$ is the vector of parameters to be estimated. \\
Here are some considerations for data generation:
The vector of true parameters is 
$\vec \theta = $ (1.6, --0.02, --0.2, --0.5, 5.4, 7.5, 1.2, 1.5). 
The age at diagnosis is uniformly distributed on intervals $[20, 40)$, $[40, 65)$ and $[65, 80]$, and the proportions of age at diagnosis coming from these three intervals are  0.25, 0.25 and 0.50 respectively; $60\%$ of cases are men; the proportions of stage I, II and III are 0.50, 0.30 and 0.20. The population hazard is assumed to follow a Weibull distribution, the scale and shape parameters being $75$ and $11$ respectively. 
The maximum follow-up time (from diagnosis) is fixed to fifteen years and the censoring rate is about $35\%$. We generate 1000 repetitions of each sample of size $n=\{250, 500,1000,2000\}$.\\
\begin{table}[!h]
\caption*{Table:}
\begin{center} 
\renewcommand{\arraystretch}{0.75}
\tabcolsep=3.5pt
\begin{tabular}{r r r r r r r r r r}
\hline
\hline 
 $n$ & ind. & $\gamma_{0}=1.6$ & $\gamma_{1}=-0.02$ & $\gamma_{2}=-0.2$ & $\gamma_{3}=-0.5$ & 
 $\beta=5.4$ & $\eta_{0}=7.5$ & $\eta_{1}=1.2$ & $\eta_{2}=1.5$ \\
 \cline{1-10}
 250 & $mean$ & 1.689 & $-0.023$ & $-0.236$ & $-0.572$ & 4.967 & 7.081 & 1.296 & 1.490   \\  
 & $sd$ & 0.191 & 0.056 & 0.175 & 0.192 & 2.044 & 3.201 & 1.209 & 1.498   \\  
 & $\overline{se^{*}}$ & 0.190 & 0.056 & 0.161 & 0.185 & 2.965 & 4.621 & 1.433 & 1.866   \\
 & $cp$ & 0.9680 & 0.966 & 0.963 & 0.967 & 0.778 & 0.804 & 0.873 & 0.931    \\
 \cline{1-10}
  500 & $mean$ & 1.647 & $-0.022$ & $-0.221$ & $-0.540$ & 5.167 & 7.259 & 1.227 & 1.504    \\  
 & $sd$ & 0.118 & 0.036 & 0.104 & 0.117 & 1.642 & 2.533 & 0.873 & 1.127   \\  
 & $\overline{se^{*}}$ & 0.124 & 0.037 & 0.104 & 0.121 & 1.994 & 3.059 & 0.929 & 1.2409   \\
 & $cp$ & 0.964 & 0.964 & 0.955 & 0.955 & 0.838 & 0.818 & 0.887 & 0.947    \\
 \cline{1-10}
 1000 & $mean$ & 1.622 & $-0.022$ & $-0.209$ & $-0.517$ & 5.317 & 7.477 & 1.238 & 1.460   \\
 & $sd$ & 0.081 & 0.025 & 0.068 & 0.079 & 1.267 & 1.994 & 0.672 & 0.819   \\
 & $\overline{se^{*}}$ & 0.083 & 0.025 & 0.070 & 0.082 & 1.365 & 2.104 & 0.649 & 0.842  \\
 & $cp$ & 0.954 & 0.945 & 0.964 & 0.960 & 0.878 & 0.891 & 0.920 & 0.944  \\ 
 \cline{1-10}
 2000 & $mean$ & 1.608 & $-0.020$ & $-0.202$ & $-0.506$ & 5.394 & 7.534 & 1.241 & 1.496   \\
 & $sd$ & 0.056 & 0.017 & 0.048 & 0.055 & 0.896 & 1.386 & 0.446 & 0.573   \\
 & $\overline{se^{*}}$ & 0.057 & 0.017 & 0.048 & 0.056 & 0.930 & 1.428 & 0.445 & 0.579   \\ 
 & $cp$ & 0.947 & 0.955 & 0.955 & 0.950 & 0.922 & 0.924 & 0.935 & 0.956   \\
 \hline
\end{tabular}
\end{center}
\end{table} 

\newpage

\noindent\textbf{Web Appendix E: Performances of the beta--TNEH model on data simulated from a Weibull mixture cure model}\\
\begin{enumerate}
\item Times to death due to the disease are generated from a Weibull mixture cure model; i.e. the net survival is written as follows:
$$S_{\rm net}(t;\vec \Phi) = \pi +(1-\pi)\exp\biggl\{-\biggl(\frac{t}{a}\biggr)^{b}\biggr\}$$
where $t$ is the time since diagnosis, $\vec \Phi = (\pi,a,b)$ is the vector of parameters with $\pi$ the cure fraction.\\
The population hazard (expected hazard) is assumed to follow a Weibull distribution, the scale and shape parameters being $75$ and $11$ respectively.
The age at diagnosis is uniformly distributed on intervals $[15, 58)$, $[58, 65)$, $[65, 70)$ and $[70, 75]$, with the same proportion (0.25) of the observations coming from each interval. The maximum follow-up time (from diagnosis) is fixed to fifteen years and we generate 1000 repetitions of sample of size 2000. 
We consider two sets of parameters corresponding to two cases:\\
\textbf{Case 1}: $\vec \Phi=(0.10,0.70,0.80)$ \\
The excess hazard is high just after the diagnosis and decreases to become very close to zero within the follow-up interval. The censoring rate is about $7\%$.\\
\textbf{Case 2}: $\vec \Phi=(0.50,1.90,0.80)$ \\ 
The excess hazard is low excess hazard just after the diagnosis and decreases but remains less close to zero than the case 1, within the follow-up interval. The censoring rate is about $26\%$.   

\item We fit the beta--TNEH model (Equation 6 in the paper) without covariate to the simulated data and assess its performances by computing the bias, the root mean square error and the coverage probability for the cure fraction, and the net survival at times $t$ = 5, 10 and 15 years.  
\end{enumerate}

\begin{figure}[!ht]
\centerline{\includegraphics[width=18cm,height=18cm]{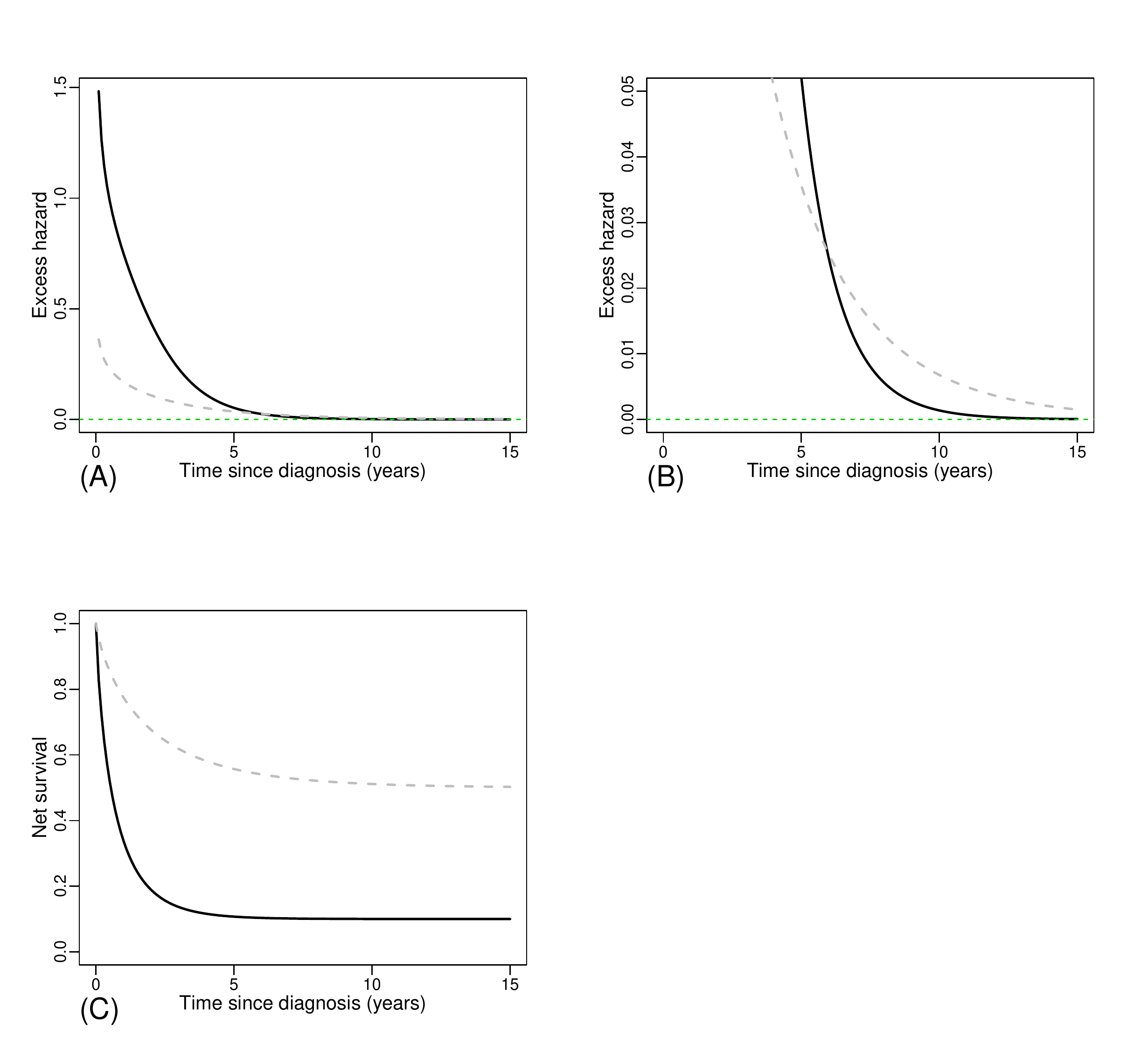}}
\caption*{Excess hazards (A), excess hazards with focus on very low values (B) and net survivals (C) functions from the Weibull mixture cure model. The vector of true parameters are $\vec \Phi=(0.10,0.70,0.80)$ and $\vec \Phi=(0.50,1.90,0.80)$ corresponding respectively to the simulation case 1 (black lines) and the simulation case 2 (dashed grey lines).}
\end{figure} 

\begin{figure}[!ht]
\centerline{\includegraphics[width=18cm,height=9cm]{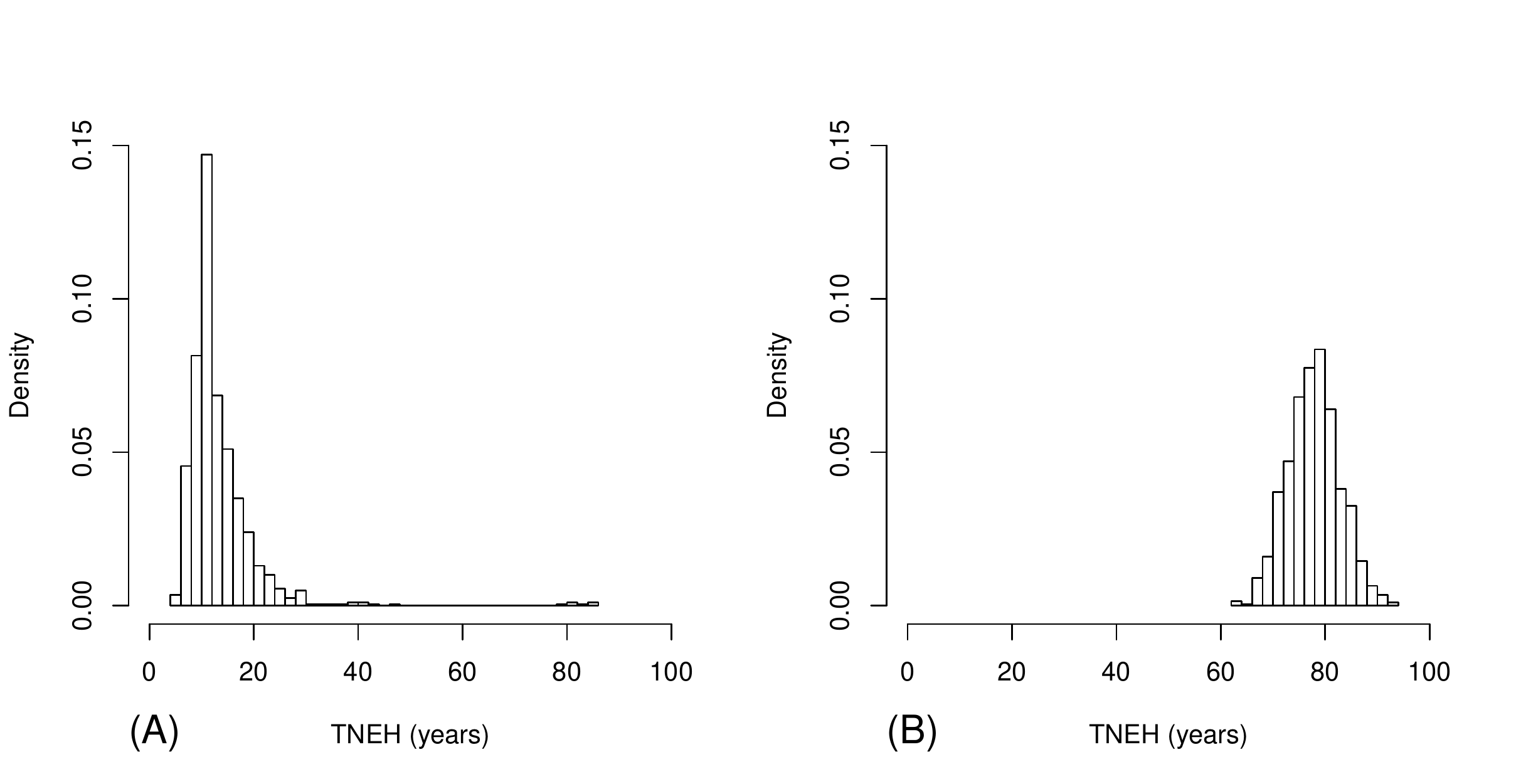}}
\caption*{Distributions of the 1000 estimates of the TNEH when the beta--TNEH model is fitted for each of the 1000 samples of size 2000 generated from the Weibull mixture cure model: (A) corresponds to Case 1 and (B) corresponds to Case 2.}
\end{figure} 

\begin{table}[!ht]
\caption*{Table: Beta--TNEH model performances when fitted to data generated from a Weibull mixture cure model (1000 repetitions of samples of size 2000): "True" is the value from the true model which generate the data, "Bias" is the mean of the biases from the beta--TNEH estimates, "RMSE" is the root mean square error of the estimates and "cp" is the coverage probability.}
\begin{center} 
\begin{tabular}{p{2cm} l r r r r}
\hline
\hline 
\textit{Case 1}   & & $S_{\rm net}(5years)$ & $S_{\rm net}(10years)$ & $S_{\rm net}(15years)$ & $Cure$ $rate$ \\
 & True & 0.107256 & 0.100203 & 0.100008 & 0.100000    \\  
 & Bias & $-0.001039$ & 0.001092 & 0.001268 & 0.001276  \\ 
 & RMSE &  0.007982 & 0.008429 & 0.008461 & 0.008463  \\ 
 & cp &  0.944000 & 0.941000 & 0.941000 & 0.941000  \\  
\hline 
\textit{Case 2}   & & $S_{\rm net}(5years)$ & $S_{\rm net}(10years)$ & $S_{\rm net}(15years)$ & $Cure$ $rate$ \\
 & True & 0.557170 & 0.511460 & 0.502697 & 0.500000    \\  
 & Bias & $-0.025030$ & 0.003064 & 0.011230 & 0.013911  \\ 
 & RMSE &  0.028640 & 0.016508 & 0.019877 & 0.021511  \\ 
 & cp &    0.583000 & 0.920000 & 0.855000 & 0.815000  \\ 
\hline 
\end{tabular}
\end{center}
\end{table}

\clearpage
\newpage

\noindent\textbf{Web Appendix F: Population mortality hazard in France}\\
Expected mortality rates were derived from the observed mortality rates available by sex, annual age, year of death (1975 to 2012), Department of residence and provided by the Institut National de la Statistique et des Etudes Economiques. For a given sex and a given Department, these observed mortality rates were smoothed for ages above 15 using a Poisson regression model that included a bidimensional smoothing spline of year and age. Mortality rates were projected for the years 2013 to 2017 using this same model. Expected mortality rates were also derived for the whole France. This work has been done by the biostatistical unit of the Hospices Civils de Lyon, using \textit{mgcv} package in R software.\\
Here for illustration purposes, we plot the population mortality hazard at years 1995, 1999, 2003, 2007, 2010 and 2013.
\begin{figure}[!h]
\centerline{\includegraphics[width=19cm,height=11cm]{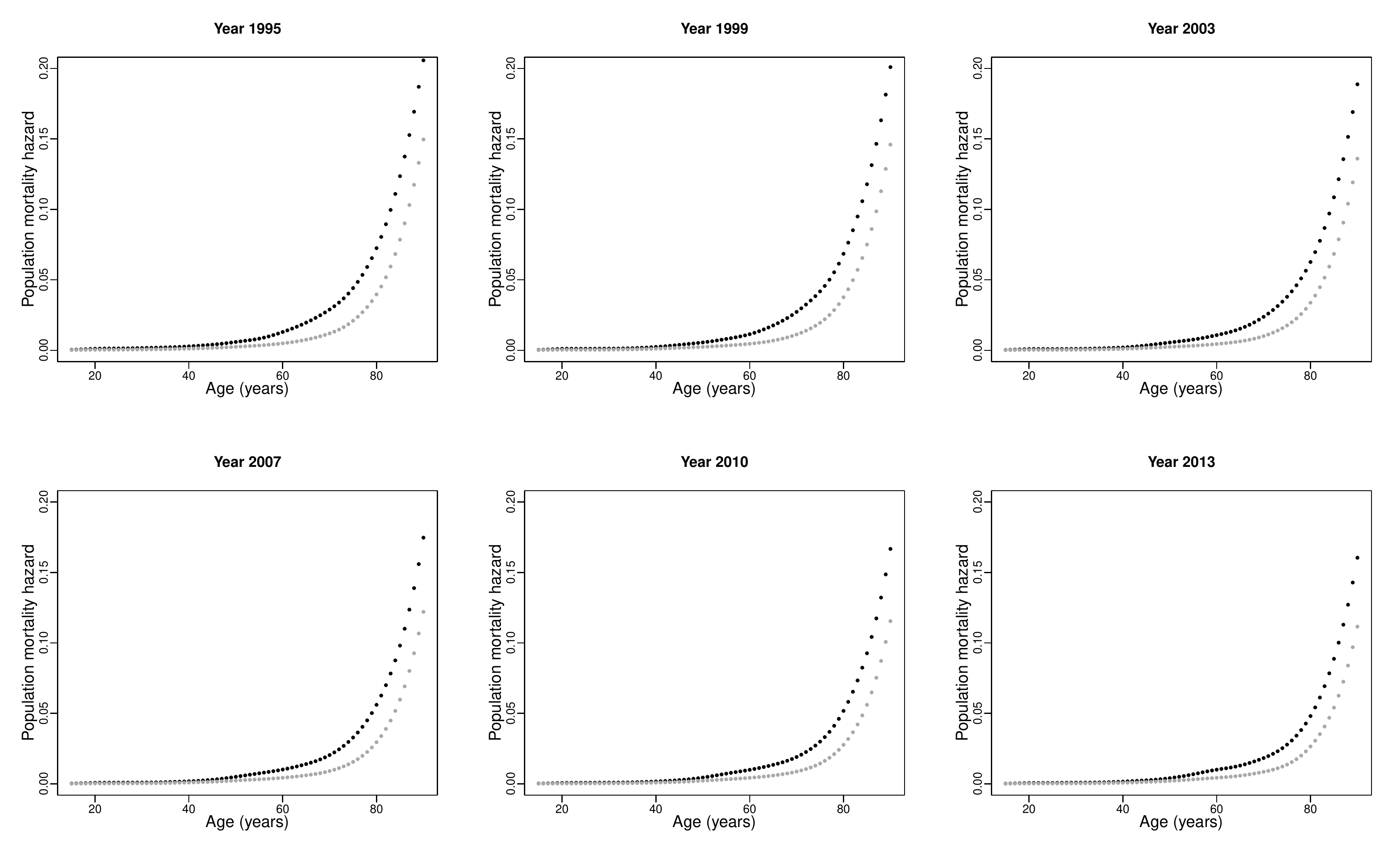}}
\caption*{French population mortality hazard in men (dark dots) and in women (grey dots).}
\end{figure}